\documentclass[pra, aps,  twocolumn,superscriptaddress,
showpacs,nofootinbib]{revtex4-2}

\usepackage[colorlinks,urlcolor=blue,citecolor=blue,linkcolor=blue]{hyperref}

\usepackage{physics}
\usepackage[english]{babel} 
\usepackage[english]{layout}
\usepackage{amsmath,amsfonts,amssymb}
\usepackage{graphicx}
\usepackage{float}

\usepackage{color}
\usepackage{braket}
\usepackage{version}
\usepackage{array,multirow,makecell}
\usepackage{afterpage}
\usepackage[toc,page]{appendix} 
\usepackage{bbold}

\renewcommand{\k}{{\bf k}}
\newcommand{\p}{{\bf p}}
\newcommand{\x}{{\bf x}}
\newcommand{\q}{{\bf q}}
\newcommand{\up}{\uparrow}
\newcommand{\dn}{\downarrow}

\newcommand{\down}{\downarrow}

\renewcommand{\k}{{\bf k}}

\newcommand{\R}{{\bf R}}

\newcommand{\son}{|1\rangle}
\newcommand{\stw}{|2\rangle}
\newcommand{\sth}{|3\rangle}

\newcommand{\beq}{\begin{equation}}
\newcommand{\eeq}{\end{equation}}

\newcommand{\bl}[1]{{\color{black}#1}}

\begin{document}

\date{\today}

\title{Hybrid-pair superfluidity in a strongly driven Fermi gas}

\author{Brendan C. Mulkerin}
\thanks{These two authors contributed equally to this work.}
\affiliation{School of Physics and Astronomy, Monash University, Victoria 3800, Australia}

\author{Olivier Bleu}
\thanks{These two authors contributed equally to this work.}
\affiliation{School of Physics and Astronomy, Monash University, Victoria 3800, Australia}

\author{Cesar R. Cabrera}
\affiliation{Institute for Quantum Physics, Universität Hamburg, Luruper Chaussee 149, 22761 Hamburg, Germany}
\affiliation{The Hamburg Centre for Ultrafast Imaging, Universität Hamburg, Luruper Chaussee 149, 22761 Hamburg, Germany}

\author{Meera M. Parish}
\affiliation{School of Physics and Astronomy, Monash University, Victoria 3800, Australia}

\author{Jesper Levinsen}
\affiliation{School of Physics and Astronomy, Monash University, Victoria 3800, Australia}

\begin{abstract}
We explore the paired superfluid phases of a %
Fermi gas in the presence of a continuous Rabi drive. We focus on the case where two components are \bl{Rabi coupled strongly enough that they form}
hybrid superpositions, \bl{which in turn interact} %
with an uncoupled third component. Using a generalized Bardeen-Cooper-Schrieffer (BCS) ansatz, we show that there are two coupled superfluid order parameters, and we obtain the associated free energy and quasiparticle excitation spectrum. We find that we can drive BCS-BCS, BCS-Bose-Einstein condensate (BEC) and BEC-BEC crossovers purely by varying the detuning of the Rabi drive from the bare transition, with the precise crossover depending on the sign of the underlying interactions between the coupled and uncoupled components. We furthermore identify an exotic excited branch which features both normal to BCS superfluid transitions, as well as a BCS-BEC-BCS crossover. Introducing a generalized Thouless criterion, we show that this behavior is reflected in the critical temperature for superfluidity. Our Rabi-coupled scenario also possesses additional thermodynamic properties related to the pseudospin of the coupled components, which provide novel signatures of the state of the many-body system. The Rabi-driven %
Fermi gas thus emerges as a unique platform for engineering and probing a rich array of multi-band superfluid phases.
\end{abstract}

\maketitle

\section{Introduction}

The exquisite experimental control of ultracold atomic gases has enabled a remarkably accurate description of superfluid Fermi gases. As a result, the Bose-Einstein condensate to Bardeen-Cooper-Schrieffer (BEC-BCS) crossover has emerged as a paradigmatic many-body system~\cite{Giorgini2008,Bloch2008, Bloch2012}, which in turn aids our understanding of other less tunable strongly correlated systems such as nuclear matter \cite{Hen2014,Ohashi2020} and potentially even high-$T_c$ superconductors \cite{loktev2001,chen2005}. Key to this progress is the fact that the underlying interactions can be manipulated with magnetic field Feshbach resonances~\cite{Chin2010}, and that the thermodynamic properties and quasiparticle excitations can be examined by driving transitions between internal atomic states, typically using radiofrequency (rf) pulses~\cite{Chin2004}.

Recently, it has become possible to fundamentally modify the behavior of the quantum gas itself using a continuous rf driving field. In the context of two-component Bose gases, this has, for instance, led to the observation of Rabi oscillations and classical bifurcations \cite{Matthews1999,Zibold2010,Cominotti2023,Recati2022}, the modification of the effective interactions \cite{Nicklas2011,Nicklas2015,Shibata2019,Sanz2022} and beyond mean-field effects \cite{Lavoine2021,Hammond2022}, as well as the realization of analogs of ferromagnetic phase transitions \cite{Cominotti2023,Zenesini2024}. For Fermi gases, the focus has primarily been on the so-called Fermi polaron problem~\cite{Massignan2014,Scazza2022}, where the Rabi oscillations between two internal states of an impurity atom immersed in a fermionic medium have been found to be sensitive to the polaron quasiparticle properties~\cite{kohstall2012,Koschorreck2012,Parish2016,Scazza2017,Oppong2019,Adlong2020,Vivanco2023,Hu2023Rabi,mulkerin2024,Wasak2024}. In particular, the recent experiment~\cite{Vivanco2023} has demonstrated that even the quasiparticles themselves can be strongly modified by an applied rf field.

\begin{figure}[t]
\includegraphics[width=.95\columnwidth]{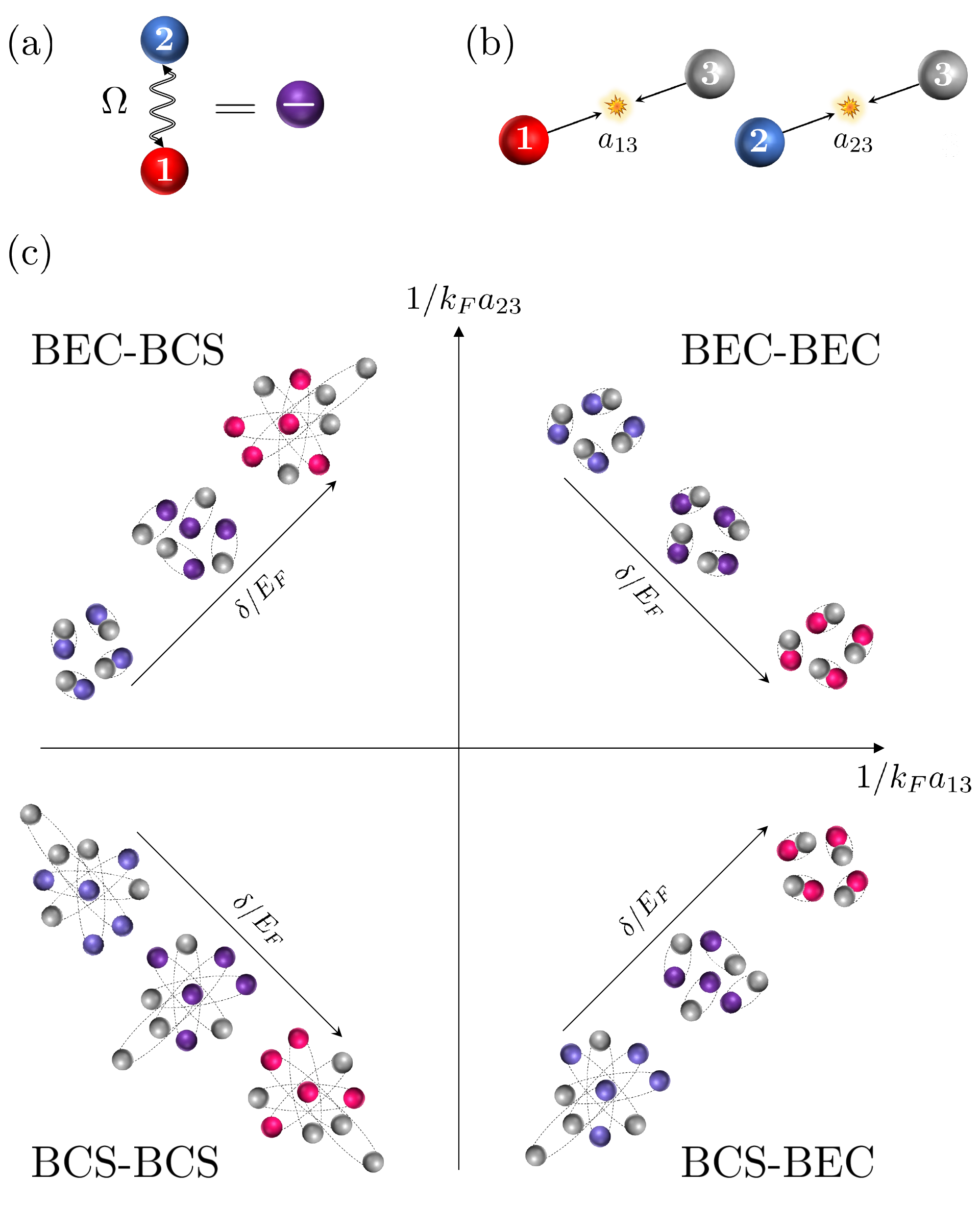}
\caption{(a) Rabi-dressed atom (purple) that is obtained by driving the transition between state $\ket{1}$ (red) and state $\ket{2}$ (blue). (b) Both states $\ket{1}$ and $\ket{2}$ can interact with a third component (gray) via short range interactions that are characterized by the scattering lengths $a_{13}$ and $a_{23}$, respectively.
(c) Illustration of the four different crossovers that can be accessed by the ground state of the Rabi driven Fermi gas when varying the detuning $\delta$, depending on the signs of the scattering lengths. The hues of the Rabi-dressed atoms indicate their spin composition.}
\label{fig:schematic}
\end{figure}

Motivated by these developments, here we investigate the superfluid properties of a strongly interacting, strongly Rabi-coupled Fermi gas. The central idea is illustrated in Fig.~\ref{fig:schematic}. A \bl{continuous} rf field creates a superposition (purple) of two atomic states ($\son$=red and $\stw$=blue), where each state interacts differently with a third atomic species ($\sth$=gray). Taking the total number of atoms in the Rabi-driven components equal to the third component, i.e., $N_1+N_2=N_3$, we find that there is the prospect of driving a rich array of different superfluid crossovers---BCS-BCS, BCS-BEC, and BEC-BEC---purely by varying the detuning of the rf field from the bare transition. While our proposed setup naturally involves three states, we emphasize that for sufficiently large Rabi coupling, \bl{where its strength exceeds the Fermi energy, the system} effectively becomes a two-component Fermi gas where the upper Rabi-dressed state is minimally occupied. In this sense, our focus is distinct from previous theory works on three-component Fermi gases, such as those exploring %
analogs of color superfluidity~\cite{He2006,Paananen2006,Zhai2007,Abuki2007,Catelani2008,Ozawa2010,Nishida2012,Kurkcuoglu2018}.

Our results are based on a generalized BCS variational ansatz, which gives us access to the free energy, the two superfluid order parameters (one for each interaction channel), the quasiparticle excitation spectrum, and thermodynamic properties. The ground state of the free energy displays the various superfluid crossovers illustrated in Fig.~\ref{fig:schematic}(c) as a function of the rf detuning. However, in addition, we find an excited-state solution, corresponding to a saddle point of the free energy, where it is possible to achieve both a normal-BCS superfluid transition, as well as a BCS-BEC-BCS crossover. This excited state is topologically distinct from the ground state and can possibly be accessed dynamically. Calculating the critical temperature for fermion pairing by using the Thouless criterion, we find that the ground state is always connected to a pairing instability at finite temperature, while the excited branch can exhibit a vanishing critical temperature, reflecting the transition to a normal state. 

From the free energy, we can also access other thermodynamic variables such as the chemical potential and the magnetization of the Rabi-coupled species. The latter is particularly interesting, as it provides an experimental probe of the many-body system which is unique to the Rabi-coupled scenario, being absent in the conventional BCS-BEC crossover. We furthermore calculate the probability of having three particles at short range (this can be thought of as a ``disconnected'' three-body contact), which is proportional to the three-body loss rate, and we discuss the requirements for this to remain small throughout the crossover.

The manuscript is organized as follows. In Sec.~\ref{sec:model} we describe our model Hamiltonian of the three-component Rabi-driven Fermi gas, and discuss the resulting hybrid superpositions of the Rabi-coupled components. In Sec.~\ref{sec:BCS_theory} we introduce a generalized BCS ansatz and derive the free energy, gap equations, number equations, and quasiparticle excitations. In Sec.~\ref{Sec:thouless} we use the Thouless criterion to evaluate the superfluid critical temperature of the ground and excited state solutions, and in Sec.~\ref{sec:results} we examine the associated range of possible superfluid crossovers at zero temperature. Finally, in Sec.~\ref{sec:conc} we conclude and provide a brief outlook. Further details of our calculations are provided in the appendices, and a detailed discussion of the two-body properties in the Rabi-coupled Fermi gas is provided in our related paper~\cite{Shortpaper}.

\section{Model} \label{sec:model}

We consider a three-component Fermi gas where two of the components are coupled by a strong Rabi drive. The system is described by the total Hamiltonian $\hat{H}= \hat{H}_0 + \hat{V}_1+\hat{V}_2$ with
\begin{align} \label{eq:Ham0} 
 \hat{H}_0&=\sum_{\k, j}\epsilon_{\k  j} \hat{f}_{\k j}^\dagger \hat{f}_{\k j}+ \frac{\Omega}{2} \sum_\k (\hat{f}_{\k 1}^\dagger \hat{f}_{\k 2}+h.c.),
\end{align}
and the interaction terms
\begin{align} \label{eq:V} 
  \hat{V}_j&= g_{j3} \sum_{\k,\k',\q} \hat{f}_{\k+\q j}^\dagger \hat{f}^{\phantom{\dagger}}_{\k  j}\hat{f}_{\k'-\q 3}^\dagger  \hat{f}^{\phantom{\dagger}}_{\k' 3} .
\end{align}
Here, $\hat{f}_{\k j}$ ($\hat{f}_{\k j}^{\dagger}$) are annihilation (creation) operators for fermions of species $j=\{1,2,3\}$. The states $|1\rangle$ and $|2\rangle$ are coupled via a Rabi drive \bl{which we treat within the rotating wave approximation. The drive has} strength $\Omega$  \bl{and it} is detuned by $\delta$ with respect to the bare $|1\rangle$-$|2\rangle$ transition. We therefore take the dispersions for each species at momentum $\k$ to be $\epsilon_{\k 1 }\equiv \epsilon_{\k  }= k^2/2m$, $\epsilon_{\k 2 } =\epsilon_\k+\delta$ and $\epsilon_{\k 3 }= k^2/2m_3$. We take a general mass ratio $m_3/m$ for species $\sth$ compared with the coupled species $\son$ and $\stw$, since many of our results for the paired superfluid phases only depend on the reduced mass $m_r=m m_3/(m+m_3)$. They thus apply to general mass-imbalanced Fermi mixtures~\cite{zaccanti2023massimbalancedfermimixturesresonant}, as well as equal-mass $^6$Li or $^{40}$K gases where the three species would represent three distinct hyperfine states.
For instance, the scenario where three hyperfine components of $^6$Li are simultaneously present has already been experimentally realized in Refs.~\cite{Ottenstein2008,Huckans2009,Williams2009,Lompe2010,schumacher2023}. Throughout, we work in units where the reduced Planck constant $\hbar$, the Boltzmann constant $k_{\rm B}$, and the volume are all set to 1.

In this work, we only consider $|1\rangle$-$|3\rangle$ and $|2\rangle$-$|3\rangle$  interactions, while we assume that the $|1\rangle$-$|2\rangle$ interactions are negligible, a scenario which can be achieved in $^6$Li \cite{Li2024}. To model the interactions, we use a short-range contact potential leading to 
Eq.~\eqref{eq:V} in momentum space. The bare interaction strengths $g_{j3}$ between species $\ket{j}$ and $|3\rangle$  are related to the scattering lengths $a_{j3}$ via the renormalization condition
\begin{align} \label{eq:renorm}
\frac{1}{g_{j3}} = \frac{m_r}{2\pi a_{j3}} - \sum_{\k}^{\Lambda} \frac{1}{\bar\epsilon_{\k}},
\end{align}
where $\bar{\epsilon}_\k\equiv\epsilon_{\k}+\epsilon_{\k3}$. $\Lambda$ is a UV cutoff on momentum which will eventually be taken to infinity to ensure that our results are fully renormalized and independent of the specific UV physics.

In the absence of interactions, the Rabi coupling dresses the $\son$-$\stw$ states and the non-interacting Hamiltonian can be diagonalized as
\begin{align}\label{eq:ham0v2}
 \hat{H}_0&=\sum_{\k \pm}(\epsilon_{\k }+\epsilon_\pm)\hat{f}_{\k \pm}^\dagger \hat{f}_{\k \pm}+\sum_{\k }\epsilon_{\k 3 } \hat{f}_{\k 3}^\dagger \hat{f}_{\k 3},
\end{align}
where we have introduced the dressed fermion operators
\begin{align} \label{eq:dressedops}
\begin{pmatrix}
  \hat{f}_{\k-}\\
 \hat{f}_{ \k +}
 \end{pmatrix}&=\begin{pmatrix}
  c &&-s\\
s &&  c \end{pmatrix} \begin{pmatrix}
   \hat{f}_{\k 1}\\
   \hat{f}_{\k 2}
 \end{pmatrix},
\end{align}
and associated transformation coefficients 
\begin{alignat}{1} \
c^2 &= \frac{1}{2}\left(1 + \frac{\delta}{\sqrt{\Omega^2 +\delta^2} }  \right), \nonumber \\
s^2 &= \frac{1}{2}\left(1 - \frac{\delta}{\sqrt{\Omega^2 +\delta^2} }  \right), \nonumber \\ \label{eq:coeffs}
cs &= \frac{1}{2}\frac{\Omega}{\sqrt{\Omega^2 +\delta^2} },
\end{alignat}
with $c^2+s^2=1$. We see that $c^2$ and $s^2$ correspond to the $\son$ fractions of the lower and upper dressed states, respectively (and vice versa for $\stw$). The dressed fermions' kinetic energies are now two split parabolas, $\epsilon_{\k }+\epsilon_\pm$, with
\begin{align}\label{eq:rabidressed}
\epsilon_{\pm} =\frac{1}{2}\left(\delta\pm \sqrt{\delta^2+\Omega^2}\right).
\end{align}

\section{BCS mean field theory}
\label{sec:BCS_theory}

In the following, we wish to describe the many-body system in the presence of both interactions and Rabi coupling. To this end, it is most convenient to use the grand canonical Hamiltonian
\begin{align}
\hat{K}= \hat{H}-\sum_j \mu_j \hat{N}_j,
\end{align}
where $\mu_j$ is the chemical potential for each state $|j\rangle$. Due to the presence of the Rabi drive, the Hamiltonian does not commute with the number operators $\hat{N}_j=\sum_{\k }\hat{f}_{\k j}^\dagger \hat{f}_{\k j}$ for species $|1\rangle$ and $|2\rangle$, and instead we have 
\begin{align}
 [\hat{H},\hat{N}_3]=0 , ~~ [\hat{H},\hat{N}_1+\hat{N}_2]=0. 
\end{align}
Thus, in the following we characterize the system by the two chemical potentials $\mu_1=\mu_2=\mu$ and $\mu_3$.

\subsection{Generalized BCS ansatz}

To investigate the ground state in the configuration with balanced densities $\langle \hat{N}_3\rangle=\langle \hat{N}_1+\hat{N}_2\rangle$, we use a generalized version of the BCS variational ansatz that accounts for the two possible pairing channels \cite{Parish2005,acton2005}
\begin{align}
    \ket{\psi}&= \prod_\k(u_\k+ v_{\k}  \hat{f}_{\k 3}^\dagger \hat{f}_{-\k a}^\dagger)\ket{0}, \label{eq:ansatz}
\end{align}
where  $u_\k$ and $v_\k$ are variational coefficients satisfying $|u_\k|^2+|v_\k|^2=1$ as in the standard BCS theory.
In writing this ansatz, we have introduced the Fermi operator $\hat{f}_{\k a}$ which is related to the operators of species $\ket{1}$ and $\ket{2}$ via the transformation
\begin{align}\label{eq:matopb}
\bl{\begin{pmatrix}
  \hat{f}_{\k a}\\
 \hat{f}_{ \k s}
 \end{pmatrix}=\begin{pmatrix}
  c_\k &&-s_\k\\
s_\k^* &&  c_\k^* \end{pmatrix} \begin{pmatrix}
   \hat{f}_{\k 1}\\
   \hat{f}_{\k 2}
 \end{pmatrix},}
\end{align}
with $|c_\k|^2+|s_\k|^2=1$.
Note that\bl{, unlike the exact coefficients $c$ and $s$ in Eq.~\eqref{eq:dressedops},} $c_\k$, $s_\k$ are variational coefficients that define the distribution between the two pairing states and thus remain to be determined\footnote{In general, once the system is interacting we have $\hat{f}_{\k a/s}\neq \hat{f}_{\k \mp }$. \bl{Here, we use the indices $a,s$ by analogy with antisymmetric and symmetric superpositions.}}.

Taking the expectation value, $\mathcal{F}=\langle\hat{K}\rangle$, we obtain the free energy in the BCS state\footnote{We note that there are two extra contributions to the free energy $\mathcal{F}$ in Eq.~\eqref{eq:expectHam13&23}, namely $g_{13} \sum_{\k\k'} |c_\k|^2 |v_\k|^2|v_{\k'}|^2$ and $g_{23} \sum_{\k\k'}  |s_\k|^2 |v_\k|^2|v_{\k'}|^2$. However, these vanish upon renormalization of the interaction, i.e., when we take $\Lambda\rightarrow\infty$ and $g_{j3}\rightarrow0$, and therefore we drop these terms in the following. 
}
\begin{alignat}{1} 
\label{eq:expectHam13&23} 
 \mathcal{F}  = &\sum_{\k}\zeta_\k|v_\k|^2
+g_{13} \sum_{\k\k'} c_\k u_\k^* v_\k  c_{\k'}^* u_{\k'} v_{\k'}^*  \nonumber \\
&+g_{23} \sum_{\k\k'} s_\k u_\k^* v_\k  s_{\k'}^* u_{\k'} v_{\k'}^*  
 ,
\end{alignat}
where we have introduced
\begin{align}\label{eq:zetak}
\zeta_\k= \xi_{\k 3 }+ \xi_{\k } +\delta |s_\k|^2
-\frac{\Omega}{2} (c_\k s_\k^*+c_\k^* s_\k) ,
\end{align}
with $\xi_{\k 3 }=\epsilon_{\k 3 }-\mu_3$ and $\xi_{\k  }=\epsilon_{\k  }-\mu$. 

\subsection{Gap equations}

We now minimize the free energy with respect to the variational parameters $(u_\k,v_\k,c_\k,s_\k)$. Without loss of generality, we take these to be real. Then, taking advantage of the normalization conditions and using the parametrization $u_\k=\cos \theta_\k$, $v_\k=\sin \theta_\k$, $c_\k=\cos \phi_\k$, $s_\k=\sin \phi_\k$, the minimization conditions $\partial_{\theta_\k} \mathcal{F}=0$ and $\partial_{\phi_\k} \mathcal{F}=0$ give
 \begin{subequations} \label{eq:minthetaphi13&23}
\begin{align}\label{eq:mintheta1323}
0&= \zeta_\k u_\k v_\k +  (u_\k^2-v_\k^2) \left( \Delta_{1}  c_\k - \Delta_{2}  s_\k \right),
\\ 
0&= \left[2\delta c_\k s_\k -\Omega  (c_\k^2-s_\k^2)\right]v_\k^2
-2 u_\k v_\k \left(\Delta_{1}  s_\k + \Delta_{2}  c_\k\right)   . \label{eq:minphi1323}
\end{align}
 \end{subequations}
Here, we have introduced the two \bl{real} order parameters $\Delta_{1}$ and  $\Delta_{2}$ defined by
 \begin{subequations} \label{eq:gapdef1323}
\begin{align}
\Delta_{1}&=g_{13}\sum_{\k}\langle \hat{f}_{-\k1} \hat{f}_{\k3}\rangle= g_{13} \sum_{\k} c_{\k} u_{\k} v_{\k},\\
\Delta_{2}&=g_{23}\sum_{\k}\langle \hat{f}_{-\k2} \hat{f}_{\k3}\rangle= -g_{23} \sum_{\k} s_{\k} u_{\k} v_{\k}.
\end{align}
\end{subequations} 
\bl{Note that, had we considered arbitrary phases of the Rabi drive and variational parameters, the phase difference between $\Delta_1$ and $\Delta_2$ would be locked to that of the Rabi drive, up to multiples of $\pi$. This would not change any of our conclusions below, and hence we keep all of these parameters real.}

Following the usual BCS theory and solving Eq.~\eqref{eq:mintheta1323} for $\theta_\k$, we obtain $\tan{2\theta_\k}=-2 (\Delta_{1}c_\k -\Delta_{2}s_\k)/\zeta_\k,$
from which we can deduce
\begin{align}\label{eq:ukvk1323}
&v_\k^2= \frac{1}{2}\left(1-\frac{\zeta_\k}{2E_\k}\right),~~
u_\k v_\k= -\frac{\Delta_{1} c_\k- \Delta_{2} s_\k}{2E_\k},
\end{align}
 with
\begin{align}
&E_\k= \sqrt{\zeta_\k^2/4+ (\Delta_{1} c_\k- \Delta_{2} s_\k)^2}. \label{eq:Ek1323}
\end{align}
Then, combining Eqs.~\eqref{eq:gapdef1323} and \eqref{eq:ukvk1323}, we get two coupled gap equations
\begin{subequations} \label{eq:gapBCS1323}
\begin{align}
\frac{\Delta_{1}}{g_{13}}&=- \sum_\k \frac{\Delta_{1} c_\k^2-\Delta_{2} c_\k s_\k}{2E_\k},\\
\frac{\Delta_{2}}{g_{23}}&=- \sum_\k \frac{\Delta_{2} s_\k^2-\Delta_{1} c_\k s_\k }{2E_\k}.
\end{align}
 \end{subequations}
Using Eq.~\eqref{eq:renorm} then results in the fully renormalized form
 \begin{subequations} \label{eq:gapBCS1323v2}
\begin{align}
\label{eq:gapBCS1323_a}
\frac{m_r}{2\pi a_{13}}\Delta_{1}&=- \sum_\k \left[\frac{\Delta_{1} c_\k^2-\Delta_{2} c_\k s_\k}{2E_\k} - \frac{\Delta_1}{\bar{\epsilon}_{\k}} \right ],\\
\frac{m_r}{2\pi a_{23}}\Delta_{2}&=- \sum_\k \left[ \frac{\Delta_{2} s_\k^2-\Delta_{1} c_\k s_\k }{2E_\k} - \frac{\Delta_2}{\bar{\epsilon}_{\k}} \right ].
\end{align}
 \end{subequations}

It is also useful to explicitly evaluate the free energy along the contour specified by the minimization with respect to $\theta_\k$. To this end, injecting Eq.~\eqref{eq:ukvk1323} into \eqref{eq:expectHam13&23}, one can obtain
\begin{align} 
\mathcal{F}_{\text{BCS}} &=- \frac{\Delta_{1}^2}{g_{13}}- \frac{\Delta_{2}^2}{g_{23}}+ \sum_{\k}\left(\frac{\zeta_\k}{2}-E_\k\right). 
\end{align}
Again, utilizing Eq.~\eqref{eq:renorm} results in the fully renormalized form
\begin{align} 
\mathcal{F}_{\text{BCS}} =&- \frac{m_r\Delta_{1}^2}{2\pi a_{13}}- \frac{m_r\Delta_{2}^2}{2\pi a_{23}} 
+ \sum_{\k}\left(\frac{\zeta_\k}{2}-E_\k +\frac{\Delta_1^2+\Delta_2^2}{\bar{\epsilon}_{\k}}\right)
. \label{eq:minGP1323} 
\end{align}
We see that we recover the gap equations by taking $\partial_{\Delta_{i}}{\mathcal{F}_{\text{BCS}}}=0$. 

Although $\mathcal{F}_{\text{BCS}}$ and the gap equations derived above have a similar form to those in the standard BCS theory~\cite{Parish2014}, here our equations involve the additional variational coefficients $c_\k$ and $s_\k$. To calculate these coefficients, one needs to use Eq.~\eqref{eq:minphi1323}. Injecting Eq.~\eqref{eq:ukvk1323} in Eq.~\eqref{eq:minphi1323}, it can be rearranged as an equation for the ratio $t_\k\equiv s_\k/c_\k$:
\begin{align} \nonumber
& \left[t_\k \left(2 \frac{\bar{\xi}_\k  +\delta}{\Omega}-\frac{\Delta_{2}}{\Delta_{1}}\right)+2\frac{\Delta_{2} \bar{\xi}_\k}{\Delta_{1}\Omega}-1\right]
\\ 
&~~~~~~~~\times \frac{ \left(t_\k^2  +2t_\k\frac{\delta}{\Omega} -1\right)}{4\left(t_\k+\frac{\Delta_{2}}{\Delta_{1}}\right)^2 \left(1-t_\k\frac{\Delta_{2}}{\Delta_{1}}\right)}=\frac{\Delta_{1}^2}{\Omega^2}, \label{eq:tk1323}
\end{align}
where we define $\bar{\xi}_\k\equiv \xi_\k+\xi_{\k3}$.

It is insightful to take a look at the limit of large $\k$ where the right hand side is negligible. In that case, the term in the first bracket gives the approximate solution
\begin{align} \label{eq:approxtk1323}
t_\k&\simeq\frac{1-2\frac{\bar{\xi}_\k}{\Omega}\frac{\Delta_{2} }{\Delta_{1}}}{ 2 \frac{\bar{\xi}_\k  +\delta}{\Omega}-\frac{\Delta_{2}}{\Delta_{1}}}.
\end{align}
Thus, we see that $t_\k$ remains finite as $k\rightarrow\infty$, and we have $t_\infty= - \Delta_{2}/\Delta_{1}$, which is the correct asymptotic behavior in order for the gap equations \eqref{eq:gapBCS1323} to be renormalizable. We also note that Eq.~\eqref{eq:approxtk1323} becomes the exact solution in the low-density limit (see  Appendix~\ref{app:low_dens}).

Since Eq.~\eqref{eq:tk1323} is a cubic equation in $t_\k$ it has, in general, three solutions. However, for a given momentum and choice of parameters, only one of these satisfies the correct asymptotic behavior $t_\infty= - \Delta_{2}/\Delta_{1}$ needed for renormalization. Although the analytic form is cumbersome, it is possible to explicitly construct the function $t_\k$ that corresponds to the real solution by combining the analytical solutions of the cubic equation with Heaviside functions. This function can then be used when solving the gap equations (and the number equation~\eqref{eq:numBCS} below).

In general, we find that the gap equations can admit two types of nontrivial solutions with $\Delta_{1,2}\neq 0$ depending on the relative sign of the two order parameters. Although in different contexts, we note that similar in-phase or out-of-phase solutions have also been reported in models of two-band superconductors \cite{Suhl1959,Leggett66} and superfluidity mediated via orbital Feshbach resonances \cite{Iskin2016,He2016,Laird2020} which also feature two coupled gap equations. 

\begin{figure}
\includegraphics[width=\columnwidth]{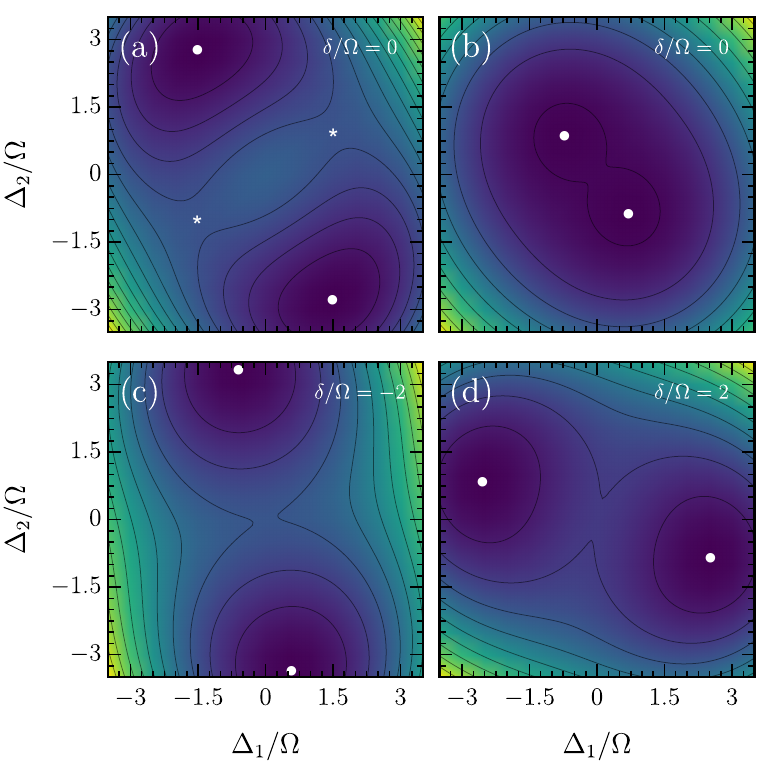}
\caption{Colormaps of the zero-temperature free energy \eqref{eq:minGP1323} as a function of the two order parameters.  (a,b) The free energy at zero detunings for $(a_{13}^{-1},a_{23}^{-1})/\sqrt{2 m_r \Omega}=(1.1,1.3)$, and $(a_{13}^{-1},a_{23}^{-1})/\sqrt{2 m_r \Omega}=(0.55,0.65)$, respectively.  (c,d) Free energy for  $(a_{13}^{-1},a_{23}^{-1})/\sqrt{2 m_r \Omega}=(1.1,1.3)$ at $\delta/\Omega=\mp 2$. In all panels, we fixed the chemical potentials to be $(\mu+\mu_3-\epsilon_{-})/2=0.25\Omega$. The white dots highlight the minima, while the white stars in panel (a) show the saddle points.}
\label{fig:FreeE}
\end{figure}

In our model \bl{where we take $\Omega>0$}, the solution with $\text{sgn}(\Delta_{2}/\Delta_{1})=-1$ corresponds to the stable ground state, while the one with $\text{sgn}(\Delta_{2}/\Delta_{1})=+1$ corresponds to an excited state, which appears as a saddle point of the free energy. This is illustrated in Fig.~\ref{fig:FreeE}, which shows colormaps of $\mathcal{F_\text{BCS}}(\Delta_1,\Delta_2)$. Indeed, we can observe in panel (a) the existence of 4 stationary points corresponding to a pair of minima with $\text{sgn}(\Delta_{2}/\Delta_{1})=-1$, and a pair of saddle points appearing in the orthogonal direction in the $(\Delta_1,\Delta_2)$ plane. We see that the excited branch is topologically distinct from the ground state, in the sense that one of the order parameters must go through zero when evolving from one state to another. 
\bl{Furthermore, as is evident in Fig.~\ref{fig:FreeE}, the free energy is very flat in the region around the saddle point.} %
This suggests the possibility of tuning into a many-body excited state that is long lived compared to the typical timescales of the system. 

We also find that the saddle points do not always exist and can disappear when the scattering lengths are modified as shown in Fig.~\ref{fig:FreeE}(b) or by changing the detuning $\delta/\Omega$ as shown in Fig.~\ref{fig:FreeE}(c,d). We  observe that the minima can move within their quadrants when the scattering lengths and/or detuning are modified. In the following, we will see that this allows access to different kinds of crossovers as a function of the detuning.

We also observe that the two types of BCS solutions are connected to the two-body bound states in the limit of vanishing density (see Appendix~\ref{app:low_dens}). However, we stress that while it has been shown that the existence of two bound states requires both scattering lengths to be positive~\cite{Shortpaper}, we find that this is not a necessary condition for two solutions of the order parameters to exist in the many-body regime. In Section \ref{Sec:thouless}, we discuss how one can use a generalized Thouless criterion to investigate the existence of the excited state in the many-body regime.

\subsection{Quasiparticle excitations} \label{sec:QPergs}

We also calculate the quasiparticle energies, i.e., the energies of the single-particle excitations on top of the generalized BCS ansatz in Eq.~\eqref{eq:ansatz}. To do so, we introduce the Bogoliubov operators
 \begin{subequations}  \label{eq:trans2}
\begin{align}
\hat{\gamma}_{\k\up}&=u_\k \hat{f}_{\k3}-v_\k \hat{f}^{\dagger}_{-\k a},
\\
\hat{\gamma}_{-\k\dn}&=v_\k \hat{f}^{\dagger}_{\k3}+u_\k \hat{f}_{-\k a},
\end{align}
 \end{subequations}
which have the property $\hat{\gamma}_{\k\sigma}\ket{\psi}=0$, where $\sigma\in\{\up,\dn\}$, so that the BCS state acts as the vacuum of Bogoliubov excitations.
To determine the quasiparticle energies we first calculate the following expectation values \cite{ring2004nuclear} 
\begin{subequations}  
\begin{align}
E_{\k\sigma}&=\langle\hat{\gamma}_{\k\sigma}\hat{H}\hat{\gamma}^{\dagger}_{\k\sigma}\rangle-\langle\hat{H}\rangle,\\
E_{\k s}&=\langle\hat{f}_{\k s}\hat{H}\hat{f}^{\dagger}_{\k s}\rangle-\langle\hat{H}\rangle,
\\
g_{\k\dn s}&=\langle\hat{\gamma}_{\k\dn}\hat{H}\hat{f}^{\dagger}_{\k s}\rangle=\langle\hat{f}_{\k s}\hat{H}\hat{\gamma}^{\dagger}_{\k \dn}\rangle,\\
g_{\k\up s}&=\langle\hat{\gamma}_{\k\up}\hat{H}\hat{f}^{\dagger}_{\k s}\rangle=\langle\hat{f}_{\k s}\hat{H}\hat{\gamma}^{\dagger}_{\k \up}\rangle.
\end{align}
\end{subequations}
We find
\begin{subequations} \label{eq.qenergies1323}
\begin{align} \label{eq.Eup1323}
E_{\k\up}&=E_{\k}+\frac{h_\k}{2},\\
E_{\k\dn}&=E_{\k}-\frac{h_\k}{2},\\
E_{\k s}&= \xi_{\k  } +\delta c_\k^2 +  \Omega  c_\k s_\k ,\\ 
g_{\k\dn s}&=- v_\k \left[s_\k  \Delta_{1} + c_\k  \Delta_{2}\right]- u_\k \left[c_\k s_\k \delta -\frac{\Omega}{2}(c_\k^2-s_\k^2)\right],\\
g_{\k\up s}&= 0,
\end{align}
\end{subequations}
where we have introduced
\begin{align}
h_{\k}&=\xi_{\k 3}-\xi_{\k }-\delta s_\k^2 +\Omega c_\k s_\k.
\end{align}

While the set of operators $\{\hat{f}^{\dagger}_{\k s},\hat{\gamma}^{\dagger}_{\k \dn},\hat{\gamma}_{\k\up}^\dag\}$ span the possible quasiparticle operators, we can see that $\hat{f}^{\dagger}_{\k s}$ and $\hat{\gamma}^{\dagger}_{\k \dn}$ are coupled via $g_{\k\dn s}$. On the other hand, $\hat{\gamma}_{\k\up}^\dag$ remains uncoupled. Thus, diagonalizing we find that the three quasiparticle energies are given by $E_{\k\up}$ and $E_{\k\pm}$ with
\begin{align} \label{eq.Epm1323}
E_{\k\pm}&=\frac{1}{2}\left(E_{\k s}+E_{\k \dn}\pm \sqrt{(E_{\k s}-E_{\k \dn})^2+4 g_{\k\dn s}^2}\right).
\end{align}
The $\pm$ quasiparticles are hybridized between the states $   \hat{\gamma}_{\k\dn}$ and $ \hat{f}_{\k s}$, i.e.,
\begin{align}\label{eq:trans3}
\begin{pmatrix}
   \hat{\gamma}_{\k-}\\
   \hat{\gamma}_{\k+}
 \end{pmatrix}&=\begin{pmatrix}
  \mathcal{C}_\k&&-\mathcal{S}_\k\\
 \mathcal{S}_\k &&    \mathcal{C}_\k \end{pmatrix} \begin{pmatrix}
   \hat{\gamma}_{\k\dn}\\
   \hat{f}_{\k s}
 \end{pmatrix},
\end{align}
with the respective fractions given by
\begin{subequations}
\begin{align}
   \mathcal{C}_\k^2 &=\frac{1}{2}\left(1+\frac{E_{\k s}-E_{\k \dn}}{E_{\k +}-E_{\k -} }\right),\\
      \mathcal{S}_\k^2 &=\frac{1}{2}\left(1-\frac{E_{\k s}-E_{\k \dn}}{E_{\k +}-E_{\k -} }\right).
\end{align}
\end{subequations}

\subsection{Number equations}
To relate the chemical potentials to the particle densities $n_1$, $n_2$, and $n_3$, the gap equations should be supplemented with the number equations
\begin{subequations}
\begin{align}
    n_1+n_2 &= -\pdv{\mathcal{F}}{\mu}, \\
    n_3 & = - \pdv{\mathcal{F}}{\mu_3} \, .
\end{align}
\end{subequations}
Note that the densities $n_1$ and $n_2$ cannot be fixed individually since they are coupled by the Rabi drive. However, one can determine the different components via the Rabi-drive parameters, i.e., $n_2=\partial_{\delta}{\mathcal{F}}$. This gives us access to an effective spin degree of freedom, or magnetization, as we discuss in Sec.~\ref{sec:results}.

In the case of the unpolarized superfluid phase at zero temperature, the free energy \eqref{eq:minGP1323} yields
\begin{align} \label{eq:numBCS}
n_3=n_1+n_2&=\frac{1}{2}
\sum_\k\left(1-\frac{\zeta_\k}{2E_\k}\right) \, ,
\end{align}
These expressions can also be arrived at directly from the BCS ansatz \eqref{eq:ansatz} by taking expectation values of the number operators. 

In the following, we find it useful to introduce a Fermi wave vector $k_F$ %
and an associated effective Fermi energy $E_F$, defined by
\begin{align}
    k_F=(6\pi^2 n_3)^{1/3}, \qquad E_{F}=\frac{k_F^2}{4m_r}.
\end{align}
We define $E_F$ in this way so that our results for the zero-temperature paired superfluid are independent of mass ratio. Note that $E_F$ reduces to the usual definition for $m=m_3$.

\section{Thouless criterion} \label{Sec:thouless}
We now include the effect of a finite temperature $T$ at the level of the BCS theory. This allows us to evaluate the critical temperature within the so-called Thouless criterion \cite{Thouless1960}. Strictly speaking, the Thouless criterion determines the temperature $T^*$ at which the system is unstable towards forming pairs, which is a prerequisite for superfluidity rather than superfluidity itself. It is known to provide a reasonable estimate for the superfluid critical temperature $T_c$ in the weakly interacting BCS regime, where pair-breaking excitations dominate; however, it overestimates $T_c$ when the ansatz yields tightly bound pairs.

The effect of thermal fluctuations is related to the thermal occupation of the quasiparticle states. This can be incorporated into the theory by adding an additional term in the free energy
 \begin{align} 
\mathcal{F}_{\text{tot}} &=\mathcal{F}_{\text{BCS}}-\frac{1}{\beta}\sum_{\k j}   \ln(1+e^{-\beta E_{\k j}}),
 \label{eq:minGP1323Th} 
\end{align}
where $\beta \equiv 1/T$ %
and 
$E_{\k j}$ corresponds to the three quasiparticle energies which are given by Eqs.~\eqref{eq.Eup1323} and \eqref{eq.Epm1323}.
Finite temperature also modifies the number equations that relate the chemical potentials to the densities. In particular, the densities of the Rabi-coupled Fermi gas in the normal phase correspond to
\begin{subequations} \label{eq:num-norm}
\begin{align}
    n_1 + n_2 = & \sum_\q \left( \frac{1}{e^{\beta (\xi_{\q}+\epsilon_{+})}+1} + \frac{1}{e^{\beta (\xi_{\q}+\epsilon_{-})}+1} \right), \\
    n_3 = & \sum_\q \frac{1}{e^{\beta \xi_{\q3}}+1} \, .
\end{align}
\end{subequations}

We wish to investigate the phase transition between the normal and superfluid states. To do so, one can expand the free energy around the limit of vanishing $\Delta_{1,2}$. We find (see Appendix~\ref{sec:freeEexpansion} for details) that it takes the form
\begin{align} \nonumber
\mathcal{F}_{\text{tot}} &\simeq -\frac{1}{\beta}\sum_{\k} \ln(1+e^{-\beta\xi_{\k3}}) -\frac{1}{\beta} \sum_{\k \pm}\ln (1+e^{-\beta(\xi_\k+\epsilon_{\pm})})   \\ 
&~~- \left(\Delta_{1}\quad \Delta_{2}\right) \mathbf{T}^{\text{mb}}(0)^{-1} \begin{pmatrix}
    \Delta_{1}\\ \Delta_{2}
\end{pmatrix} + \mathcal{O}(\Delta_{i}^2\Delta_{j}^2).
\label{eq:Ftotexpansionfinal}
\end{align}
Here, the first line corresponds to the free energy of the non-interacting Rabi-coupled Fermi gas, while the second line encodes the effect of pairing.  $\mathbf{T}^{\text{mb}}$ denotes the many-body $T$ matrix of the Rabi-coupled Fermi gas, which reads
\begin{align}\label{eq:manybT}
\bold{T}^{\text{mb}}(\omega)^{-1}=\begin{pmatrix}
    \frac{1}{g_{13}} &0 \\ 0 &  \frac{1}{g_{23}}
\end{pmatrix}-\begin{pmatrix}
    \Pi_{11}^{\text{mb}}(\omega) &\Pi_{12}^{\text{mb}}(\omega)  \\ \Pi_{21}^{\text{mb}}(\omega)  &\Pi_{22}^{\text{mb}}(\omega)  
\end{pmatrix},
\end{align}
with
\begin{align}\nonumber
 \Pi_{11}^{\text{mb}}(\omega)&=\sum_\k \left[c^2\frac{1-f_{\k3}-f_{\k-}}{\omega-\bar{\xi}_{\k}-\epsilon_{-}}+s^2 \frac{1-f_{\k3}-f_{\k+}}{\omega-\bar{\xi}_{\k}-\epsilon_{+}}\right],
\\ \nonumber
 \Pi_{22}^{\text{mb}}(\omega)&=\sum_\k \left[s^2\frac{1-f_{\k3}-f_{\k-}}{\omega-\bar{\xi}_{\k}-\epsilon_{-}}+c^2 \frac{1-f_{\k3}-f_{\k+}}{\omega-\bar{\xi}_{\k}-\epsilon_{+}}\right],
\\ \nonumber
 \Pi_{12}^{\text{mb}}(\omega)&=\sum_\k cs\left[ \frac{1-f_{\k3}-f_{\k+}}{\omega-\bar{\xi}_{\k}-\epsilon_{+}}- \frac{1-f_{\k3}-f_{\k-}}{\omega-\bar{\xi}_{\k}-\epsilon_{-}}\right]
 ,
\end{align}
and $\Pi_{21}^{\text{mb}}(\omega)= \Pi_{12}^{\text{mb}}(\omega)$. Here, we have introduced the Fermi occupations of the Rabi-dressed fermions $f_{\k\pm}=1/(e^{\beta (\xi_{\k}+\epsilon_{\pm})}+1)$ and of the third component $f_{\k3}=1/(e^{\beta \xi_{\k3}}+1)$, and as before $\bar{\xi}_{\k}=\xi_{\k3}+\xi_{\k}$. The many-body $T$ matrix can be renormalized similarly to the gap equations (see Eqs.~\eqref{eq:gapBCS1323} and \eqref{eq:gapBCS1323v2}).

The expansion of the free energy in Eq.~\eqref{eq:Ftotexpansionfinal} indicates that 
the system is unstable towards forming a superfluid state with non-zero order parameters when
\begin{align}\label{eq:TCfromansatz}
\det[\mathbf{T}^{\text{mb}}(0)^{-1}]=0.
\end{align}
This corresponds to a generalized version of the Thouless criterion \cite{Thouless1960}. 
\bl{We note that a Thouless criterion given by the determinant of a two-by-two matrix also occurs in the model of two-band superfluids used in Refs.~\cite{Iskin2006,Tajima2019}. However, this model differs from the one we consider since it describes a four-component Fermi system, while we consider a system with three components.  Furthermore, in the models of two-band superfluids \cite{Suhl1959,Leggett66,Iskin2006,Tajima2019}, the interband coupling has a two-body origin and couples pairs from different bands, whereas here the Rabi coupling is a single-particle effect.}

Solving Eq.~\eqref{eq:TCfromansatz} together with the number equations in Eq.~\eqref{eq:num-norm} allows us to determine the critical temperature at fixed density. Specifically, renormalizing the diagonal elements of the inverse many-body $T$ matrix in Eq.~\eqref{eq:manybT} in the same manner as the gap equation in Eq.~\eqref{eq:gapBCS1323v2}, we find that the Thouless criterion takes the form
\begin{align}\nonumber
&\frac{m_r}{2\pi a_{13}} \frac{m_r}{2\pi a_{23}}+\left(c^2\frac{m_r}{2\pi a_{13}}+s^2 \frac{m_r}{2\pi a_{23}}\right) I_{+}
\\
&~~~~~~~+ \left(s^2\frac{m_r}{2\pi a_{13}}+c^2 \frac{m_r}{2\pi a_{23}} \right)I_{-} +I_{+}I_-=0, \label{eq:ThoulesscritII_{+}}
\end{align}
where the temperature dependence is captured in
\begin{align}
I_{\pm}=\sum_{\k}\left[ \frac{1-f_{\k\pm}-f_{\k3}}{\bar{\xi}_\k+\epsilon_{\pm}}-\frac{1}{\bar{\epsilon}_\k} \right].
\end{align}

We note that Eq.~\eqref{eq:TCfromansatz} has a richer structure than the corresponding equation in the usual scenario of a two-component Fermi gas. In particular, we find that it can have two solutions which are related to the ground and excited state solutions of the gap equations mentioned earlier. As we discuss below, unlike in the conventional two-component scenario, we also find that the Thouless criterion can be satisfied at strictly zero temperature, which is related to the fact that the excited superfluid state disappears, undergoing a transition to the normal phase at zero temperature. 

\begin{figure}
\includegraphics[width=0.94\columnwidth]{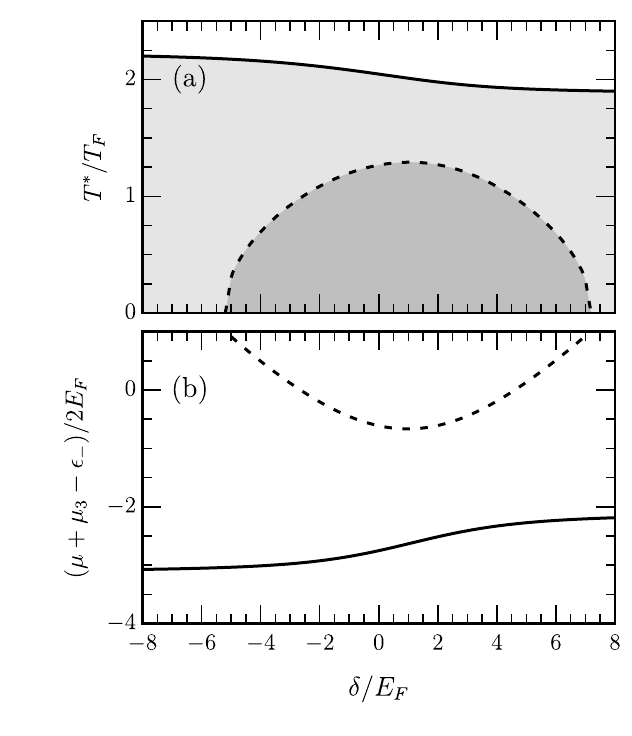}
\caption{(a) Critical temperature $T^*$ and (b) corresponding total chemical potential versus detuning for $\Omega/E_F=4$,  $(a_{13}^{-1},a_{23}^{-1})/\sqrt{2 m_r \Omega}=(1.1,1.3)$ (same scattering lengths as in Fig.~2(a,c,d)), and equal masses $m=m_3$. We show the results for the ground and excited states as solid and dashed lines, respectively.}
\label{fig:Tc2inter}
\end{figure}

Figure \ref{fig:Tc2inter} shows the critical temperature $T^*$ and total chemical potential $\mu+\mu_3$ calculated for the set of scattering lengths used in panels (a,c,d) of Fig.~\ref{fig:FreeE} at a fixed density corresponding to $\Omega/E_F = 4$. We see that for a range of detunings there are two solutions corresponding to the ground and excited states. The ground state (solid line) has the largest critical temperature and remains well defined throughout the entire range of detunings. By contrast, for the excited state (dashed line), we see that the critical temperature in Fig.~\ref{fig:Tc2inter}(a) vanishes at the same detunings where the corresponding total chemical potential (measured from the shifted continuum) reaches $2E_F$ in Fig.~\ref{fig:Tc2inter}(b). This indicates that the excited state approaches a free Fermi gas at the critical detunings. For larger magnitudes of the detuning, the excited state solution ceases to exist, i.e., there is no longer a saddle point, and we thus expect a BCS-normal state phase transition at the critical detunings.

\subsection{Thouless criterion in the limit of zero temperature}

We now wish to evaluate the generalized Thouless criterion in the limit of zero temperature and weak interactions.
To evaluate $I_{\pm}$ analytically in this limit, we assume %
a large Rabi drive $\Omega\geq E_F$. 
This implies that for weak interactions we have $\mu-\epsilon_{-}\simeq k_F^2/2m$ and $ \mu_3\simeq  k_F^2/2m_3$. In this case, $I_{-}$ becomes independent of the parameters of the Rabi drive and takes the form
  \begin{align} \nonumber
I_{-}&=\sum_{\k}\left[ \frac{ \tanh( \beta\frac{\bar{\epsilon}_{\k}-2E_F}{2} \frac{m_r}{m} )}{2(\bar{\epsilon}_\k-2E_F)}-\frac{1}{2\bar{\epsilon}_\k} \right] \\ \nonumber
&~~~~~~ + \sum_{\k}\left[ \frac{ \tanh( \beta\frac{\bar{\epsilon}_{\k}-2E_F}{2} \frac{m_r}{m_3} )}{2(\bar{\epsilon}_\k-2E_F)}-\frac{1}{2\bar{\epsilon}_\k} \right]\\ \label{eq:I_}
&\simeq
\frac{m_r k_F}{2\pi^2}\ln\left[\left(\frac{16 e^{\gamma-2} \beta E_F  }{\pi}\right)^2 \frac{m_r^2}{m m_3}\right].
\end{align}
Here, $\gamma\simeq 0.577$ is the Euler constant, and to obtain the last line we have used the fact that $I_{-}$ diverges logarithmically as temperature is taken to zero, thus allowing us to analytically capture the leading divergence~\cite{Stoof2009b}. By contrast, $I_{+}$ is finite as $T\to0$ and depends on the Rabi-drive parameters:
 \begin{align}\label{eq:Ip}
I_{+}&\simeq \sum_{\k}\left[ \frac{1-\Theta(k_F-k)}{\bar{\epsilon}_\k-2E_F+\epsilon_{+}-\epsilon_{-}}-\frac{1}{\bar{\epsilon}_\k} \right]\\ \nonumber
&=-\frac{m_r k_F}{\pi^2}\left[1+\sqrt{\frac{\epsilon_{+}-\epsilon_{-}}{2E_F}-1}\arctan\sqrt{\frac{\epsilon_{+}-\epsilon_{-}}{2E_F}-1}\right],
    \end{align}
where we have replaced the Fermi-Dirac distribution by the Heaviside function in order to evaluate the integral analytically.

The above results allow us to distinguish two scenarios depending on whether the Thouless criterion can be satisfied at zero temperature or not.
Indeed, according to Eq.~\eqref{eq:ThoulesscritII_{+}}, the divergence of $I_{-}$ implies that when $a_{13}$ and $a_{23}$ are nonzero, the zero-temperature Thouless criterion becomes
\begin{align} \label{eq:Thoulesscritzero}
s^2\frac{m_r}{2\pi a_{13}}+c^2 \frac{m_r}{2\pi a_{23}} +I_{+}=0.
\end{align}
Since $I_{+}<0$ is an even and concave function of $\delta$, the potential solutions of Eq.~\eqref{eq:Thoulesscritzero} will appear pairwise if $s^2/ a_{13}+c^2 /a_{23}\geq 2\pi|I_+|/m_r$ is realized at some detunings. Thus, such solutions can only occur when at least one of the scattering lengths is positive. A pair of $T=0$ solutions can be clearly seen in Fig.~\ref{fig:Tc2inter}(a), at the points where the excited state $T^*$ goes to zero. The paired solutions of Eq.~\eqref{eq:Thoulesscritzero} thus define the critical detunings in-between which the excited state exists for a given set of scattering lengths.

\subsection{Thouless criterion at finite temperature}

When Eq.~\eqref{eq:Thoulesscritzero} cannot be satisfied, there is no zero-temperature transition and instead, for weak interactions, the Thouless criterion \eqref{eq:ThoulesscritII_{+}} gives a finite critical temperature
\begin{align} \label{eq:TcBCS_Rabi} 
T^*/T_F= \frac{2m_r}{\sqrt{m m_3}} \frac{8e^{\gamma-2}}{\pi} e^{F(\delta,\Omega, E_F, a_{13},a_{23})}.
\end{align}
Here we have used the fact that the critical temperature for weak interactions is much smaller than the Fermi temperature $T_F\equiv E_F$, thus allowing us to approximate $I_-$ as in Eq.~\eqref{eq:I_}.
The dimensionless function $F$ is given by
\begin{align} 
F=  \frac{\pi}{2k_F} \frac{1+\frac{2\pi}{m_r}I_{+}(a_{13}s^2+a_{23}c^2)}{a_{13}c^2+a_{23}s^2+\frac{2\pi}{m_r}I_{+}a_{13}a_{23}}.
\end{align}
In the limit $E_F/\Omega\ll 1$, this function reduces to $F\simeq \frac{\pi}{2k_Fa_{-}}$, with 
\begin{align}\label{eq:dressedscatt}
a_{-}&= \frac{a_{13}c^2+a_{23}s^2-a_{13}a_{23}\sqrt{2 m_r(\epsilon_{+}-\epsilon_{-})}}{1-\sqrt{2 m_r(\epsilon_{+}-\epsilon_{-})} (a_{13}s^2+a_{23}c^2)}
\end{align}
the two-body scattering length between a particle of species $\sth$ and the lower dressed particle $\hat f_{\k-}$ [see Eq.~\eqref{eq:dressedops}], as introduced in Ref.~\cite{Shortpaper}. This result for the critical temperature closely matches the standard result for the critical temperature in the BCS limit of a conventional equal-mass two-component Fermi gas with $s$-wave scattering length $a_s<0$ \cite{SadeMelo1993}:
\begin{align}
T^*/T_F=  \frac{8}{\pi}e^{\gamma-2}\exp\left( -\frac{\pi}{2k_F|a_s|} \right).
\end{align}
In other words, in the case of a strong drive and weak interactions, we have an effective BCS state where the pairing is between species $\sth$ and the lower dressed superposition of species $\son$ and $\stw$.

\begin{figure}
\includegraphics[width=0.94\columnwidth]{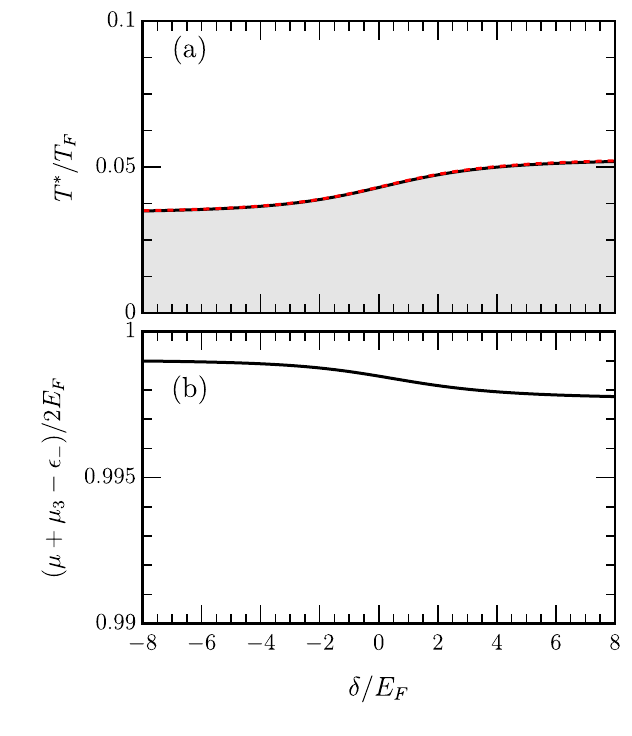}
\caption{(a) Critical temperature $T^*$ and (b) corresponding total chemical potential versus detuning for $\Omega/E_F=4$, $(a_{13}^{-1},a_{23}^{-1})/\sqrt{2 m_r \Omega}=(-1.1,-1.3)$, and equal masses $m=m_3$. The dashed red line in panel (a) shows Eq.~\eqref{eq:TcBCS_Rabi} with $F= \frac{\pi}{2k_Fa_{-}} $.}
\label{fig:Tc2interbcsbcs}
\end{figure}

Likewise, if one of the scattering lengths vanishes, i.e., if $a_{j3}\rightarrow 0$, 
the critical temperature \eqref{eq:TcBCS_Rabi} takes the form
\begin{align}\label{eq:Tcone inter}
T^*/T_F=& \frac{2m_r}{\sqrt{m m_3}} \frac{8}{\pi}e^{\gamma-2} \\ \nonumber
& \times \begin{cases} 
 \exp(\frac{\pi/c^2}{ 2k_Fa_{13}}+\frac{\pi^2}{m_r k_F}\frac{s^2}{c^2}I_{+}), &    a_{23}= 0,\\ \exp(\frac{\pi/s^2}{ 2k_Fa_{23}}+\frac{\pi^2}{m_r k_F}\frac{c^2}{s^2}I_{+}), &    a_{13}= 0. 
\end{cases}  
\end{align}
We again see that, when $s^2\rightarrow 0$ or $c^2\rightarrow 0$, respectively, the above expressions reduce to the standard expression from BCS theory.

In Fig.~\ref{fig:Tc2interbcsbcs}, we show the critical temperature and total chemical potential for a set of negative scattering lengths in the BCS limit, at fixed density $\Omega/E_F =4$. Here, we find a single solution corresponding to the ground state. We see that the critical temperature remains low but nonzero across the range of detunings, and that furthermore the numerical result for $T^*$ (solid black line) is very well approximated by Eq.~\eqref{eq:TcBCS_Rabi} with $F= \frac{\pi}{2k_Fa_{-}}$ (dashed red line). We also see that the chemical potential is close to that of free fermions.

\section{Zero-temperature ground state}
\label{sec:results}

In this section, we investigate the zero-temperature ground state of the Rabi-driven Fermi gas. We start by introducing some general properties that characterize the system besides the order parameters: namely the momentum distributions, the magnetization, and the contact parameters in section \ref{subsec:prop}. Then, in section \ref{subsec:becbcs}, we present the results for the configuration in which one of the bare interactions is negligible, $a_{23}\rightarrow 0^{-}$, and in \ref{subsec:2inter} we present the results when both $a_{13},a_{23}\neq0$. 
We will see that the first scenario allows us to probe a BEC to BCS crossover analogous to what has been probed in two-component Fermi gases using magnetic field Feshbach resonances. However, in the case of a Rabi-driven system, the entire crossover is accessible at a fixed bare scattering length by instead varying the detuning. On the other hand, we will see that in the second scenario, varying the detuning allows us to access a different kind of crossover between the two pairing channels which has no analog in the two-component Fermi gas.

\subsection{Ground-state properties}\label{subsec:prop}

\subsubsection{Density distributions and pairing correlations}

From the BCS ansatz, we can easily evaluate the density distributions and pairing correlations
 \begin{subequations} \label{eq:expectnew}
  \begin{align}
n_{\k11}&=\langle\hat{f}^{\dagger}_{\k1}\hat{f}_{\k1}\rangle=c_\k^2 v_\k^2,
\\
n_{\k 22}&=\langle\hat{f}^{\dagger}_{\k2}\hat{f}_{\k2}\rangle=s_\k^2 v_\k^2,
\\
n_{\k33}&=\langle\hat{f}^{\dagger}_{\k3}\hat{f}_{\k3}\rangle= v_\k^2,\\
n_{\k 12}&=\langle\hat{f}^{\dagger}_{\k1}\hat{f}_{\k2}\rangle=-v_\k^2 s_\k c_\k,
\\
\kappa_{\k 13}&=\langle\hat{f}_{-\k1}\hat{f}_{\k3}\rangle= u_\k v_\k c_\k,
\\
\kappa_{\k 23}&=\langle\hat{f}_{-\k2}\hat{f}_{\k3}\rangle= -u_\k v_\k s_\k.
  \end{align}
  \end{subequations}
To gain more insight, we evaluate the tails of the momentum occupations at large momenta, which, as we will see below, are related to the two-body contact parameters~\cite{Tan1,Tan2}. We find
\begin{subequations}\label{eq:tails}
   \begin{align}
   n_{\k 11}& \underset{k\rightarrow \infty}{\simeq} \frac{4m_r^2}{k^4}\Delta_{1}^2,
    \\
  n_{\k 22} & \underset{k\rightarrow \infty}{\simeq} \frac{4m_r^2}{k^4}\Delta_{2}^2,
        \\
n_{\k33}&  \underset{k\rightarrow \infty}{\simeq}  \frac{4m_r^2}{k^4}(\Delta_{1}^2+\Delta_{2}^2),
\\
   n_{\k 12} &\underset{k\rightarrow \infty}{\simeq} \frac{4m_r^2}{k^4} \Delta_{1}\Delta_{2} .
\end{align}
\end{subequations}
In a similar manner, we obtain the tails of the pairing correlations
\begin{subequations}\label{eq:tailspair}
      \begin{align}
  \kappa_{\k 13} &\underset{k\rightarrow \infty}{\simeq} -\frac{2m_r}{k^2}  \Delta_1 ,\\
          \kappa_{\k 23}&\underset{k\rightarrow \infty}{\simeq}  -\frac{2m_r}{k^2}  \Delta_2.
\end{align}
\end{subequations}
We can observe from the above expressions that we have $n_{\k12}\simeq \kappa_{\k13}\kappa_{\k23}$ at large $k$. This is not surprising since we have $n_{\k12}=-v_\k^2c_\k s_\k$ and $\kappa_{\k13}\kappa_{\k23}=-u_\k^2 v_\k^2 c_\k s_\k$, and $u_\k^2\simeq 1$ at large $k$.

We can see that the tails of the momentum distributions and pairing correlations  are directly related to the values of the order parameters $\Delta_{1,2}$, and only depend on the Rabi drive indirectly.

\subsubsection{Magnetization}
An interesting property of the present Rabi driven Fermi superfluid is its magnetization in terms of the Rabi coupled species $\ket{1}$ and $\ket{2}$. The components of the magnetization vector are given by
 \begin{subequations}
  \begin{align} \label{eq:magnetizationdef}
\mathcal{S}_z&=\frac{\sum_\k\langle \hat{f}_{\k1}^{\dagger}\hat{f}_{\k1}-\hat{f}_{\k2}^{\dagger}\hat{f}_{\k2}\rangle}{\langle \hat{N}_1+\hat{N}_2\rangle},\\
\mathcal{S}_x&=\frac{\sum_\k\langle \hat{f}_{\k1}^{\dagger}\hat{f}_{\k2}+\hat{f}_{\k2}^{\dagger}\hat{f}_{\k1}\rangle}{\langle \hat{N}_1+\hat{N}_2\rangle} .
\end{align}
      \end{subequations}
Using Eq.~\eqref{eq:expectnew}, we immediately obtain
 \begin{subequations}
  \begin{align} \label{eq:magnetization}
\mathcal{S}_z &=\frac{\sum_\k v_\k^2(c_\k^2-s_\k^2)}{\sum_\k v_\k^2},\\
\mathcal{S}_x&=-2\frac{\sum_\k v_\k^2c_\k s_\k}{\sum_\k v_\k^2} .
\end{align}
      \end{subequations}
This magnetization is affected by the interactions, whose effect is encoded in the variational coefficients $v_\k,c_\k,s_\k$, and thus it generally differs from the magnetization of the non-interacting Rabi-coupled gas.

\subsubsection{Two- and three-body contacts} 
We can also show that the tails of the momentum distributions within our model are related to the two-body contact parameters \cite{Tan1,Tan2}, which are defined as \cite{Braaten2008PRL}
\begin{align} \label{eq:2bodycontact_Def}
\mathcal{C}_{j3} &= 
(2m_rg_{j3})^2\int d^3R \langle \hat{f}_{j}^\dagger \hat{f}_{ 3}^\dagger  \hat{f}_{ 3}  \hat{f}_{ j} (\R) \rangle
\\
&= (2m_rg_{j3})^2\sum_{\k,\k'\p,\p'} \delta_{\k+\p,\k'+\p'} \langle \hat{f}_{\k j}^\dagger \hat{f}_{\p 3}^\dagger  \hat{f}_{ \p' 3} \hat{f}_{\k' j}\rangle,  \nonumber
\end{align}
where in the first line the operators are defined in real space at the same position $\R$.
Within the BCS ansatz, we find
  \begin{align} \nonumber
\langle \hat{f}_{\k 1}^\dagger \hat{f}_{\p 3}^\dagger  \hat{f}_{ \p' 3} \hat{f}_{\k' 1}\rangle&=c_\k c_{\k'}  v_\k v_{\k'} v_\p v_{\p'} \delta_{\k\k'} \delta_{\p\p'}
\\ &+c_\k c_{\k'}  v_\k v_{\k'} u_\p u_{\p'} \delta_{-\k\p} \delta_{-\k'\p'}, 
\end{align}
and
  \begin{align} \nonumber
\langle \hat{f}_{\k 2}^\dagger \hat{f}_{\p 3}^\dagger  \hat{f}_{ \p' 3} \hat{f}_{\k' 2}\rangle&=s_\k s_{\k'}  v_\k v_{\k'} v_\p v_{\p'} \delta_{\k\k'} \delta_{\p\p'}
\\ &+s_\k s_{\k'}  v_\k v_{\k'} u_\p u_{\p'} \delta_{-\k\p} \delta_{-\k'\p'}, 
\end{align}
and injecting these in Eq.~\eqref{eq:2bodycontact_Def}, we obtain 
\begin{align} 
\mathcal{C}_{j3} =4\Delta_{j}^2 m_r^2,  %
\end{align}
where we have used the definition \eqref{eq:gapdef1323} for $\Delta_{j}$, and the fact that $g_{j3}\rightarrow 0^{-}$ when we take $\Lambda\rightarrow\infty$ in Eq.~\eqref{eq:renorm}. Thus, we find that the contact parameters and the momentum distributions \eqref{eq:tails} are related as $\mathcal{C}_{j3} =\lim_{k\rightarrow \infty} k^4  n_{\k jj}$, as expected~\cite{Tan2}.

In a similar manner, we can calculate the following three-body correlation function
\begin{align} \label{eq:3bodycontact_Def}
\mathcal{G}^{(3)}_{ij3} &= 
\int d^3R \langle \hat{f}_{i}^\dagger \hat{f}_{j}^\dagger\hat{f}_{ 3}^\dagger  \hat{f}_{ 3} \hat{f}_{ j}  \hat{f}_{ i} (\R) \rangle,\\ \nonumber
 &= \sum_{\k\p\q \k'\p'\q'} \delta_{\k+\p+\q,\k'+\p'+\q'} 
 \langle \hat{f}_{\k i}^\dagger \hat{f}_{\p j}^\dagger \hat{f}_{\q 3}^\dagger \hat{f}_{\q' 3}\hat{f}_{\p' j}  \hat{f}_{\k' i} \rangle.
\end{align}
We find
  \begin{align} 
\mathcal{G}^{(3)}_{123}& =\sum_{\k\p\q } v_\k^2  v_\p^2v_\q^2 c_\k s_\p (c_\k s_\p- s_\k c_\p )\\ \nonumber
& ~~+ \sum_{\k\p\q } v_\k^2 v_{\p}u_\p v_{\q}u_\q \left(c_\k^2 s_\p s_\q+ s_\k^2 c_\p c_\q-2s_\k c_\k s_\p c_\q\right),
\end{align}
while the three-body correlation functions $\mathcal{G}^{(3)}_{113}$ and $\mathcal{G}^{(3)}_{223}$ both vanish due to Fermi statistics. Evaluating the remaining sums, and using the definition of the order parameters, we find that $\mathcal{G}^{(3)}_{123}$ can be expressed in terms of the densities and %
the order parameters as
   \begin{align} 
\mathcal{G}^{(3)}_{123}& =(n_1n_2-n_{12}^2)n_3 + \frac{\Delta_1^2}{g_{13}^2} n_2+\frac{\Delta_2^2}{g_{23}^2}n_1- 2 \frac{\Delta_1 \Delta_2 }{g_{13}g_{23}}n_{12},
\end{align}
with $n_{12}= \sum_\k n_{\k12}$.

\begin{figure}
\includegraphics[width=0.94\columnwidth]{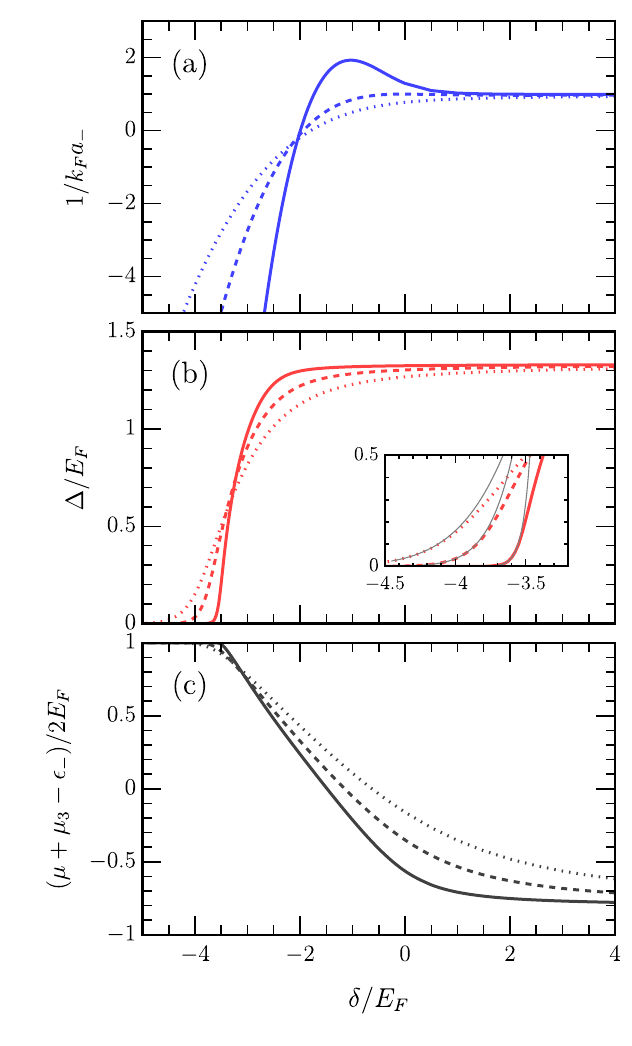}
\caption{BEC-BCS crossover with Rabi-coupled fermions. (a) Effective scattering length, (b) BCS order parameter, (c) total chemical potential measured from the (Rabi-shifted) continuum. The results are plotted for parameters $1/k_F a_{13} =1$, $a_{23}=0$, and $\Omega/E_F=1,2,3$ (solid, dashed, and dotted lines). The inset in (b) shows the analytical order parameter in the BCS regime, Eq.~\eqref{eq:analyticorder}.}
\label{fig:gap1}
\end{figure}

We then define a (disconnected) three-body contact as
\begin{align}
\mathcal{C}^{(3)}_{ij3}&=(2 m_r g_{j3})^2\mathcal{G}^{(3)}_{ij3},
\end{align}
and obtain
\begin{align} \label{eq:3bcontact}
\mathcal{C}^{(3)}_{123}& =(2m_r)^2\left(\Delta_1^2 n_2+\Delta_2^2n_1- 2 \Delta_1 \Delta_2 n_{12}\right),
\end{align}
where we have used the facts that $g_{j3}\rightarrow 0^{-}$ and $g_{13}/g_{23}\rightarrow 1$ when we take $\Lambda\rightarrow\infty$ in Eq.~\eqref{eq:renorm}. Since our ansatz only explicitly involves two-body correlations, the three-body contact is entirely determined from the two-body contacts and the densities. In this sense, the three-body contact as defined above can be considered to be ``disconnected'' and distinct from the thermodynamic variable introduced in Refs.~\cite{Braaten2011,Werner2012bosons}, which depends on genuine three-body correlations and on physics beyond the two-body scattering length (e.g., on the existence of a three-body parameter~\cite{efimov1970}.
However, the three-body contact in Eq.~\eqref{eq:3bcontact} still has an important physical interpretation as it determines the rate of three-body losses, i.e., the loss rate is directly proportional to $\mathcal{C}^{(3)}_{123}$ (see, e.g., Ref.~\cite{Laurent2017}). 

\bl{We emphasize that the physics here is distinct from the case of three-component Fermi gases featuring near-resonant interactions between all components~\cite{Nishida2012,Nishida2015,Alhyder2020}. In that case, Efimov-like trimers are expected to be strongly bound and to play a major role in both few- and many-body physics. By contrast, the fewer resonant interactions in our work suppresses Efimov physics~\cite{Naidon2017} while the strong Rabi coupling effectively reduces the system to two components. Thus %
we expect that the formation of Efimov trimers can be safely neglected in our scenario. 
}

\subsection{BEC-BCS crossover for a single interaction channel}\label{subsec:becbcs}

We now consider the configuration in which one of the bare interactions is negligible, $a_{23}\rightarrow 0^-$, a scenario that has been recently achieved in experiment~\cite{Li2024}. In this case, the corresponding pairing channel becomes irrelevant, which means that $\Delta_{2}=0$ and only the first gap equation remains, i.e., Eq.~\eqref{eq:gapBCS1323_a}. For this reason, in this subsection only we define $\Delta\equiv\Delta_1$ and we take $\Delta\geq 0$ without loss of generality.

We first wish to investigate the influence of the detuning on the system properties. To achieve this, we examine the effective scattering length $a_{-}$, which characterizes interactions between particles of species $\ket{3}$ and the lower-energy Rabi-driven superposition state of species $\ket{1}$ and $\ket{2}$ \cite{Shortpaper}. When $a_{23}=0$, the dressed scattering length \eqref{eq:dressedscatt} reduces to
\begin{align} \label{eq:aeff}
\frac{1}{a_{-}}&=\frac{1}{c^2}\left[\frac{1}{a_{13}}-s^2 \sqrt{2 m_r(\epsilon_{+}-\epsilon_{-})}\right].
\end{align}
We see that when $\delta/\Omega\gg1$, $a_{-}$ approaches the bare scattering length $a_{13}$, and that if $a_{13}>0$, $a_{-}$ can exhibit a resonance (i.e., a zero crossing of $1/a_-$) by controlling the detuning. This behavior is illustrated in Fig.~\ref{fig:gap1}(a) which shows $1/k_F a_{-}$ as a function of detuning for a fixed bare scattering length $1/k_Fa_{13}=1$ and different magnitudes of the Rabi drive: $\Omega/E_F=1$, 2, and 3. %

In Figs.~\ref{fig:gap1}(b,c) we plot the corresponding order parameters and total chemical potentials as a function of detuning obtained by solving the gap and number equations self-consistently for three values of $\Omega/E_F$. Here, we see a key result of this work: By increasing the detuning, starting from the limit $\delta/E_F\ll-1$, the ground state can be smoothly tuned from a BCS-like state, characterized by a vanishing order parameter and a chemical potential approaching  the Fermi energy shifted by the dressed continuum, to a BEC-like state with a large order parameter and $\mu + \mu_3-\epsilon_{-}<0$ at positive detunings. When $\delta/E_F\gg1$, the order parameter saturates to a constant value $\Delta\simeq 1.3 E_F$, which is independent of $\Omega/E_F$. This value is set by the bare scattering length  and corresponds to the magnitude of the order parameter in an undriven two-component Fermi gas for $k_Fa_s=1$. Finally, we can observe that as $\Omega/E_F$ increases, the crossover region broadens, and the respective BCS and BEC regimes are approached for larger negative and positive detunings, respectively. 

  \begin{figure}
    \includegraphics[width=0.94\columnwidth]{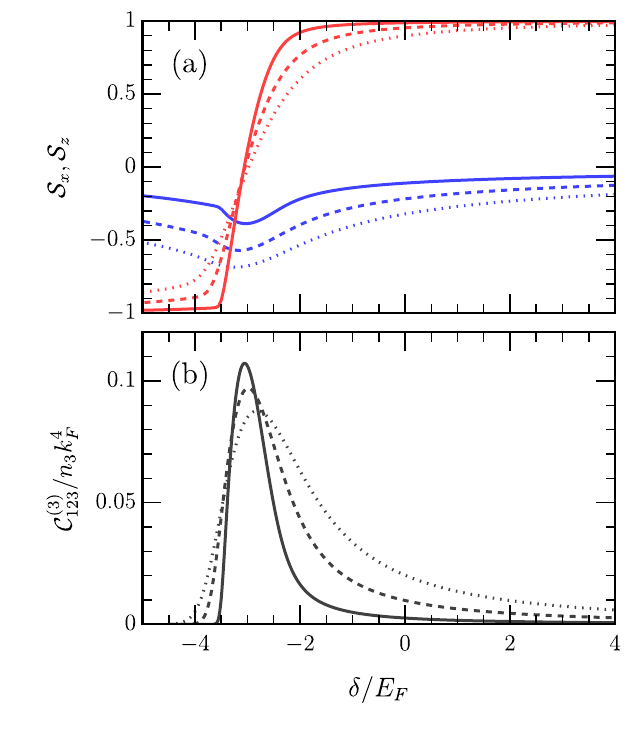}
\caption{(a) Out-of-plane and in-plane magnetization $\mathcal{S}_z$ (red) and $\mathcal{S}_x$ (blue), respectively, and (b) disconnected three-body contact. Parameters as in Fig.~\ref{fig:gap1}:  $1/k_F a_{13} =1$, $a_{23}=0$, and $\Omega/E_F=1,2,3$ (solid, dashed, and dotted lines).}
\label{fig:thermo1}
\end{figure}

The inset of Fig.~\ref{fig:gap1}(b) shows a zoomed-in view of the order parameters at negative detuning together with the analytical approximation in the BCS limit (whose derivation is presented in Appendix \ref{app:BCS_limit}):
\begin{align}
\Delta = \frac{8}{e^{2}|c|}E_{F}\,{\rm exp}\left(\frac{\pi/c^2}{2k_{F}a_{13}}-\frac{\pi^2}{m_rk_F}\frac{s^2}{c^2}I_+\right),
\end{align}
where $I_+$ is given in Eq.~\eqref{eq:Ip}. We see that the order parameter is exponentially suppressed in this limit, as also found in the standard BCS theory \cite{Stoof2009b}.
Furthermore, using Eq.~\eqref{eq:Tcone inter}, we find that the critical temperature and order parameter are related as
\begin{align} \label{eq:analyticorder}
T^*= \frac{2m_r}{\sqrt{m m_3}} \frac{e^{\gamma}}{\pi}\Delta |c| %
\end{align}
Thus, we obtain a universal relation between the critical temperature and the zero-temperature order parameter for the Rabi driven superfluid, similarly to the undriven two-component BCS superfluid \cite{Stoof2009b}. In the driven case, this relation depends on the drive parameters $\Omega$ and $\delta$ through the bare transformation coefficient $c$ (note that, likewise, a coefficient $c_\k$ multiplies the order parameter inside $E_\k$ [see Eq.~\eqref{eq:Ek1323}] which in turn appears in all three quasiparticle energies).

\begin{figure*}[th]
\includegraphics[width=\linewidth]{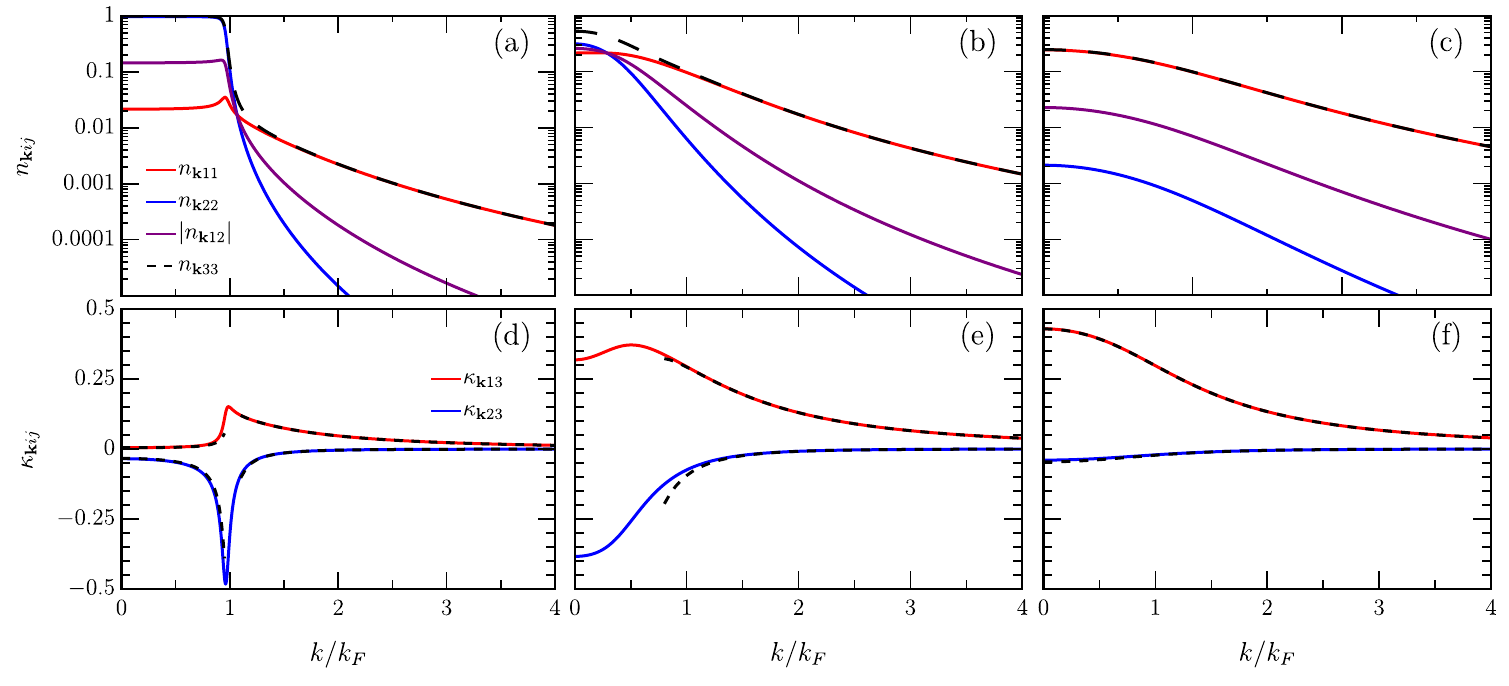}
\caption{(a,b,c)  Density distributions and (d,e,f) pairing correlations across the BEC-BCS crossover. From left to right we have $\delta/E_F=-3.4,-2.1,3$.
The dashed black lines in (d,e,f) are obtained using the low-density approximation for $c_\k$ and $s_\k$ when $\bar{\xi}_\k\geq |\epsilon_{-}|$ [Eq.~\eqref{eq:approxtk1323}] and the noninteracting coefficients [Eq.~\eqref{eq:coeffs}] when $\bar{\xi}_\k< |\epsilon_{-}|$, substituted into Eq.~\eqref{eq:expectnew}. The results are plotted for parameters 
$\Omega/E_F=1$, $1/k_F a_{13} =1$, $a_{23}=0$. 
}
\label{fig:spectra1}
\end{figure*}

We also investigate the multi-component nature of the Rabi-coupled Fermi gas, as encoded in the magnetization and three-body contact. In Fig.~\ref{fig:thermo1}(a) we see that $\mathcal{S}_z$ evolves smoothly from $-1$ to $+1$ as a function of detuning. This is natural as the ground state changes from being composed of a set of weakly perturbed Fermi seas in the $|2\rangle$ and $|3\rangle$ components at large negative detuning to being dominated by strongly bound $|1\rangle$-$|3\rangle$ pairs at positive detuning. While the zero crossing happens at zero detuning in the non-interacting system, we find that once we have interactions, $\mathcal{S}_z$ crosses zero at negative detuning roughly around the point where $\Delta$ is reduced by a factor 2 compared with the BEC limit. Furthermore, increasing $\Omega/E_F$ does not affect the position of the zero crossing, and only changes the scale of detunings over which the crossover occurs. The behavior of the zero crossing is similar to what has been obtained for the Rabi-driven Fermi polaron \cite{Vivanco2023,mulkerin2024}. The in-plane magnetization $\mathcal{S}_x$ is also similarly peaked at the point where $\mathcal{S}_z = 0$, but unlike the non-interacting case, it does not reach $-1$.

In Fig.~\ref{fig:thermo1}(b) we see that the (disconnected) three-body contact has a non-trivial behavior across the BEC-BCS crossover, and that it peaks close to the zero crossing of $\mathcal{S}_z$. This implies that the associated three-body losses are also likely to peak around this point. However, we emphasize that we would expect three-body losses to remain strongly suppressed when the three components are not all resonantly interacting. Furthermore, losses are also suppressed in the strongly driven regime as long as $\Omega\gtrsim E_F$, since in this case we effectively have a two-component Fermi gas. Indeed, this explains why $\mathcal{C}^{(3)}_{123}$ is small at positive detuning in Fig.~\ref{fig:thermo1}(b): In this regime, the two-body contact $\mathcal{C}^{(2)}_{13}$ and the associated order parameter $\Delta_1$ are large (since the $\son$-$\sth$ atoms form tightly bound pairs); however the density $n_2$ becomes negligible when $\delta\gg E_F$ and therefore the losses remain small.

\begin{figure}
    \includegraphics[width=0.94\columnwidth]{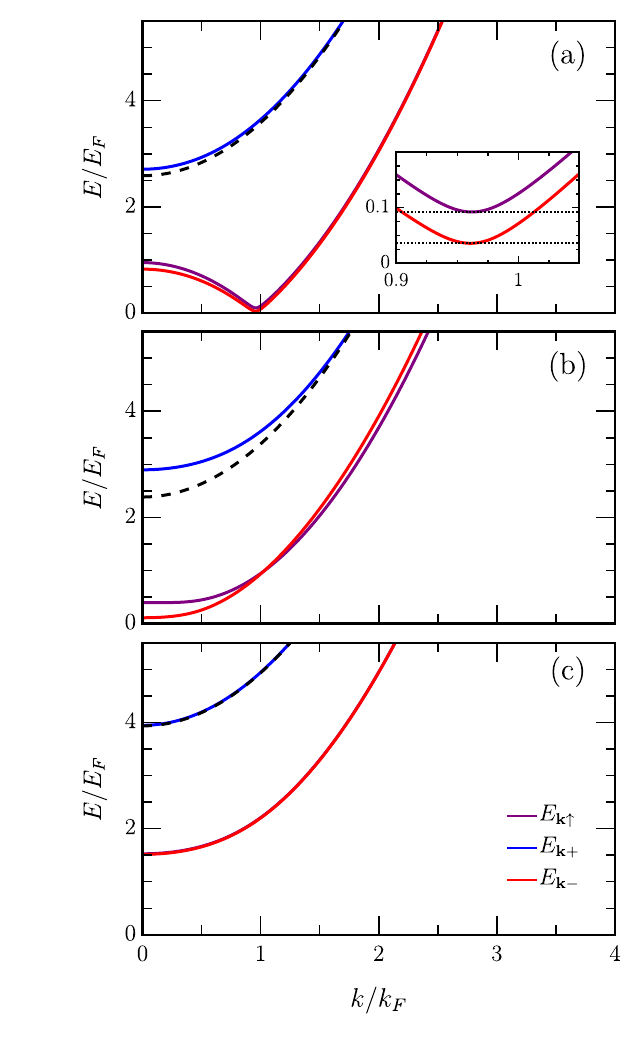}
\caption{Quasiparticle excitation spectrum $E_{\k\up}$ (purple), $E_{\k+}$ (blue), and $E_{\k-}$ (red) given in Eqs.~\eqref{eq.Eup1323} and \eqref{eq.Epm1323}, for equal masses $m=m_3$. We fix $\Omega/E_F=1$, $1/k_F a_{13} =1$ and $a_{23}=0$, and consider detunings that span the crossover, with $\delta/E_F=-3.4$ in (a), $\delta/E_F=-2.1$ in (b), and $\delta/E_F=3$ in (c).
The inset in (a) shows the zoomed in area around the minimum of the quasiparticle excitation spectrum, where the dotted lines give the magnitude of the excitation gaps derived in Eq.~\eqref{eq:minquasidisp}. The dashed black lines are the upper  non-interacting quasiparticle energies $\xi_\k+\epsilon_+$ with (a,c) $\mu-\mu_3=\epsilon_{-}$, and (b) $\mu-\mu_3=-2.6E_F$, corresponding to a lower value than $\epsilon_{-}\simeq -2.2E_F$.
}
\label{fig:quasi_ener}
\end{figure}

We now turn to show additional quantities that characterize the state of the system, namely in Fig.~\ref{fig:spectra1} the density distributions and the pairing correlations, and in Fig.~\ref{fig:quasi_ener} the quasiparticle excitation spectrum. 
In both figures, we take the bare interaction strength $k_Fa_{13}=1$, i.e., in the absence of the Rabi drive we would be on the BEC side of the bare resonance. However, according to Fig.~\ref{fig:gap1}, varying the detuning of the Rabi drive allows us to explore an effective BCS regime [$\delta/E_F=-3.4$], a crossover regime [$\delta/E_F=-2.1$], and a BEC regime [$\delta/E_F=3$].  

In Figs.~\ref{fig:spectra1}(a,b,c) we plot the density distributions $n_{\k ij}=\langle\hat{f}_{\k i}^{\dagger}\hat{f}_{\k j}\rangle$ given in Eq.~\eqref{eq:expectnew}, across three different regimes of the BEC-BCS crossover.  We see that, on the BCS side shown in panel (a), the occupation $n_{\k 33}$ interpolates between $n_{\k 11}$ and $n_{\k 22}$: For momentum $k<k_F$, $n_{\k 22}$ dominates and resembles a step function, whereas for $k>k_F$, $n_{\k 11}$ dominates. This indicates that the large-momentum pairing correlations primarily involve the interacting species $|1\rangle$, as expected. In the crossover regime shown in panel (b), we see that %
$n_{\k 22}>n_{\k 11}$ only near $k=0$, and that $n_{\k 11}$ dominates over a wider range of momentum.
On the other hand, in the BEC regime displayed in panel (c), the density distribution is primarily governed by the bound-state wave function, such that $n_{\k 11}$ is significantly larger than $n_{\k 22}$ over all momenta. 

Across all three regimes, we can see that $n_{\k22}$ and $n_{\k12}$ decay faster than $n_{\k11}$ at large momenta. Indeed, since $\Delta_2=0$, there is no contribution of order $1/k^4$ in the tails of $n_{\k22}$ and $n_{\k12}$ and the leading order contributions are instead
\begin{subequations}
   \begin{align}
     n_{\k 22} & \underset{k\to\infty}{\rightarrow} \frac{4\Delta^2 \Omega^2 m_r^4}{k^8} ,
        \\
     |n_{\k 12}| &\underset{k\to\infty}{\rightarrow} \frac{4\Delta^2|\Omega| m_r^3}{k^6} .
\end{align}
\end{subequations}

In Figs.~\ref{fig:spectra1}(d,e,f) we instead plot the pairing correlations $\kappa_{\k i3}=\langle\hat{f}_{-\k i}\hat{f}_{\k3}\rangle$. In the BCS-like limit shown in Fig.~\ref{fig:spectra1}(d), we can see that $\kappa_{\k 13}$ and $\kappa_{\k 23}$ are both substantial even though there is no direct interaction between species $\ket{2}$ and $\ket{3}$. For $k\lesssim k_F$, we have $|\kappa_{23}|>|\kappa_{13}|$ with both exhibiting peaks upon approaching the edge of the Fermi sea when $k\simeq k_F$, similar to the conventional Cooper pairing in the BCS limit of the two-component Fermi gas. On the other hand, for $k\gtrsim k_F$, $\kappa_{\k 23}$ decays %
faster than $\kappa_{\k 13}$. %
Here again, this behavior is expected since $\kappa_{\k 13}\sim 1/k^2$ at large momenta [see Eq.~\eqref{eq:tailspair}] whereas, since $\Delta_2=0$, 
the leading-order contribution to the tail of $\kappa_{\k 23}$ is
      \begin{align}
           |\kappa_{\k 23}|& \underset{k\to\infty}{\rightarrow} 2 \Delta |\Omega|\frac{m_r^2}{k^4}.
\end{align}
In the crossover regime in panel (e) we see that $\kappa_{\k 13}$ has a peak inside the Fermi sea, near $k\simeq0.5k_F$, similar to the standard BCS theory in a two-component Fermi gas. On the other hand, $\kappa_{\k 23}$ has a maximum at $k=0$, and shows no effect of Pauli blocking. This is similar to both $\kappa_{\k 13}$ and $\kappa_{\k 23}$ in the BEC-like regime shown in panel (f). 

Figure \ref{fig:quasi_ener} shows the quasiparticle energies $E_{\k\uparrow}$, $E_{\k+}$, and $E_{\k-}$ derived in Section~\ref{sec:QPergs}.
In contrast to the occupations and pairing correlations, the quasiparticle energies depend not only on the total chemical potential $\mu+\mu_3$, but also on the difference $\mu-\mu_3$. Moreover, the stability of the BCS state requires positive quasiparticle energies and this is typically satisfied for a range of  $\mu-\mu_3$. In the BCS limit, this range is set by $\Delta$ and is around the value $\mu=\mu_3+\epsilon_{-}$, the condition for an unpolarized normal gas, while in the BEC limit, the chemical potential difference must exceed the two-body binding energy in order to have a gapless superfluid. This behavior is analogous to what happens in the conventional two-component Fermi gas \cite{Parish2007}.

The BCS-like regime is shown in Fig.~\ref{fig:quasi_ener}(a). Here, as one might expect, we see that the upper branch $E_{\k+}$ is close to and slightly above the upper of the non-interacting Rabi-coupled quasiparticle branches $\xi_\k+\epsilon_{+}\simeq\epsilon_\k-E_F+\epsilon_{+}-\epsilon_{-}$ (black dashed) and that it has a minimum at $k=0$. In other words, 
the upper quasiparticle branch is only minimally involved in the pairing 
when $\Omega\gtrsim E_F$. Conversely, the lower quasiparticle branches display distinct BCS characteristics, with the energies $E_{\k\uparrow}$ and $E_{\k-}$ exhibiting gapped minima near $k\simeq k_F$. The magnitude of the excitation gaps can be calculated analytically in this limit, yielding
\begin{align} \label{eq:minquasidisp}
\text{min}\left(E_{\k \up,-}\right)\simeq|\Delta c|
\pm \frac{\Delta^2s^2}{2(\epsilon_{+}-\epsilon_{-})},
\end{align}
which matches the behavior in the inset of panel (a).
In particular, we observe a splitting of the two minima that is reminiscent of a Zeeman splitting due to an applied magnetic field.

As we move into the crossover regime of stronger interactions, the induced Zeeman splitting for $\Omega/E_F = 1$ is large enough to cause parts of the lowest quasiparticle dispersion to become negative if we take $\mu-\mu_3=\epsilon_-$. This implies that interactions can lead to an effective spin polarization in our Rabi-coupled system. Thus, in order to have a stable unpolarized superfluid, we require $\mu-\mu_3$ to be lower than $\epsilon_{-}$ such that it counteracts the Zeeman shift, as is the case for the crossover regime shown in panel (b). 
Here, $E_{\k+}$ and $E_{\k-}$ both exhibit a minimum at $k=0$, while $E_{\k\up}$ has a (shallow)  minimum at $k\simeq0.1k_F$, where the precise position depends on the choice of chemical potentials. %
Notably, we find that $E_{\k+}$ lies well above the non-interacting quasiparticle dispersion, $\xi_\k+\epsilon_{+}$. %
This indicates that pairing significantly involves both Rabi-dressed single-particle states in the crossover regime, which is consistent with the enhanced three-body contact at this detuning [Fig.~\ref{fig:thermo1}(b)].

Finally, in panel (c) the system has crossed over to the BEC regime. Here,  the quasiparticle dispersions are all clearly gapped and have minima at $k=0$. The upper quasiparticle branch  $E_{\k+}$ is essentially the same as the non-interacting quasiparticle, indicating that interactions are not affecting the upper quasiparticle branch. 
When $\delta/\Omega\gg1$, we find that the gaps at $k=0$ can be estimated as
\begin{subequations}
\begin{align} \label{eq:minquasidispBEC}
\text{min}\left(E_{\k +}\right)&\simeq\delta-\mu +\frac{\Omega^2}{4\delta}, \\ 
\text{min}\left(E_{\k \up,-}\right)&\simeq\sqrt{\tilde{\mu}^2+\Delta^2}\pm\frac{1}{2}\left(\mu-\mu_3+\frac{\Omega^2}{4\delta}\right),
\end{align}
\end{subequations}
where we have defined $\tilde{\mu}=(\mu+\mu_3-\epsilon_{-})/2$. Here, we take the chemical potential difference $\mu-\mu_3=\epsilon_-$, hence, we see the splitting of $E_{\k\up}$ and $E_{\k-}$ is small, and the energies appear degenerate. %

We also briefly comment on the other crossovers that one can obtain with a single interaction channel. First, suppose we have $k_Fa_{13}<-1$. In this case of negative scattering length, the Rabi coupling acts to make the interactions weaker, and therefore the superfluid ground state has BCS character for all detunings. Likewise, if we start with unitarity-limited interactions, $1/k_Fa_{13}=0$, we cannot induce a bound state via Rabi coupling, and as a function of detuning we have a BCS to unitary Fermi gas superfluid crossover. In Fig.~\ref{fig:schematic} these cases are obtained by taking $a_{23}\to0^-.$

\subsection{Finite interactions in both channels}\label{subsec:2inter}

\begin{figure}
\includegraphics[width=0.94\columnwidth]{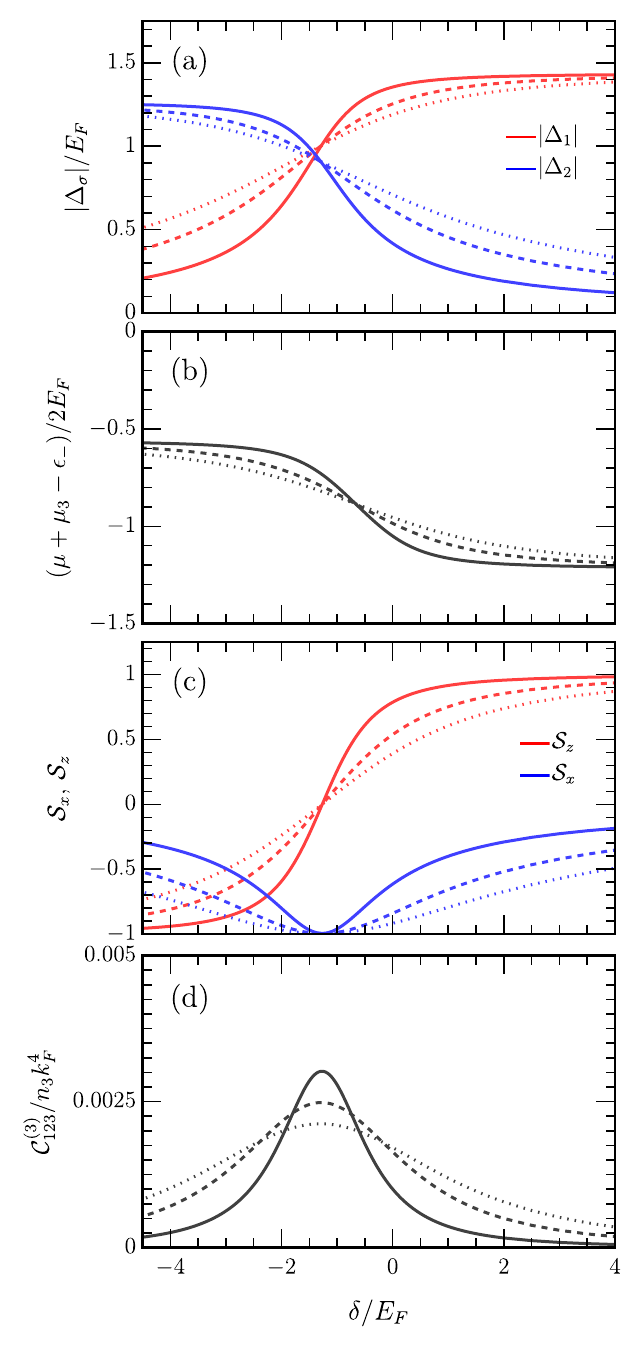}
\caption{Ground-state BEC-BEC crossover with Rabi-coupled fermions that both interact with species $\sth$. (a) Order parameters, (b) total chemical potential  with respect to the lower dressed energy, $(%
\mu + \mu_3-\epsilon_-)/2E_F$, (c) out-of-plane and in-plane magnetization $\mathcal{S}_z$ and $\mathcal{S}_x$, respectively, and (d) three-body contact. Parameters: $k_F(a_{13},a_{23})=\sqrt{2}(0.6,0.8)$ and $\Omega/E_F=1,2$, and 3 (solid, dashed, and dotted lines).}
\label{fig:becbec_gs}
\end{figure}

We now turn to the more general case where we have finite $\son$-$\sth$ and $\stw$-$\sth$ interactions. This scenario is illustrated in Fig.~\ref{fig:schematic}: If both interactions are negative and not too large, we can drive a BCS-BCS crossover by changing the detuning, where the spin character of the Rabi-driven state changes accordingly. If one interaction is positive and one negative then we get a BCS-BEC crossover which is qualitatively similar to the case of a single positive interaction. The most intriguing case is that of two positive interactions. In this regime, we have shown in Ref.~\cite{Shortpaper} that there is a range of detunings for which there exist two bound states in the two-body limit, and that these are hybrid states that continuously change their spin composition as a function of detuning. Whereas the lowest bound state exists for all detunings, the excited state unbinds at critical negative and positive detunings. Similarly, as discussed above, the many-body system can also support both a ground-state and an excited-state solution, although the critical detunings are different from the two-body case, and the excited many-body state can even exist when one of the scattering lengths is negative. We now present the properties of both of these many-body states.

\begin{figure*}
\includegraphics[width=\linewidth]{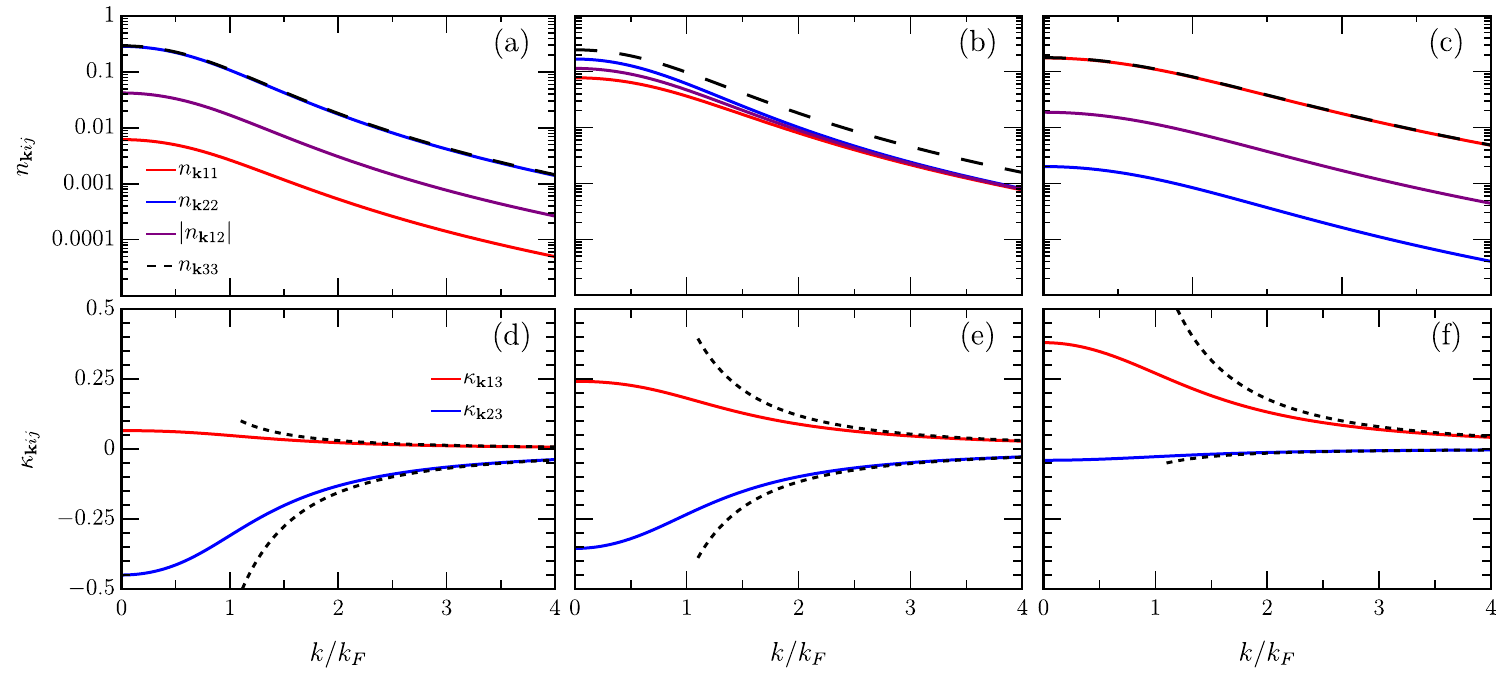}
\caption{(a,b,c) Density distributions and (d,e,f) pairing correlations across the BEC-BEC crossover. From left to right we have $\delta/E_F=-4,-1.375, 4$.
The dashed lines in (d,e,f) show the large-$k$ asymptotes [Eq.~\ref{eq:tailspair}]. Parameters: $k_F(a_{13},a_{23})=\sqrt{2}(0.6,0.8)$ and $\Omega/E_F=1$.}
\label{fig:becbecocc}
\end{figure*}

\subsubsection{Ground-state crossover}

We first look at the ground state where, unlike in the single-interaction case considered above, the detuning now drives a BEC-BEC crossover. To be concrete, in the following we consider the case $k_F(a_{13},a_{23})=\sqrt{2}(0.6,0.8)$. In Fig.~\ref{fig:becbec_gs}(a) we plot the order parameters $\Delta_{1}/E_F$ and $\Delta_{2}/E_F$ as a function of detuning for three fixed Rabi drives $\Omega/E_F=1,2,$ and 3. We see that, at large negative detuning, species $\son$ is detuned away and the order parameter $\Delta_1$ is suppressed; the system is approaching the bare BEC regime for an undriven two-component Fermi gas, i.e., with $k_Fa_s\equiv k_Fa_{23}=0.8\sqrt{2}$ as $\delta\rightarrow-\infty$. In the opposite limit for large positive  detuning, species $\stw$ is suppressed and the system is approaching the bare BEC regime for an undriven two-component Fermi gas with scattering length $k_Fa_s\equiv k_Fa_{13}=0.6\sqrt{2}$. At intermediate detuning where $\delta\simeq-1.375E_F$, the two order parameters are similar in magnitude $|\Delta_1|\simeq|\Delta_2|\simeq0.96E_F$, marking a smooth crossover between the two BEC regimes. The effect of increasing the Rabi drive is to broaden the BEC-BEC crossover, such that the bare regimes are reached at larger absolute detuning, as also found above in the case of a single interaction.

Figure \ref{fig:becbec_gs}(b) shows the total chemical potential with respect to the lower dressed energy $\epsilon_-$, i.e., $\mu + \mu_3-\epsilon_-$, plotted as a function of detuning. The total chemical potential smoothly interpolates between two distinct BEC regimes: For large negative detuning, the chemical potential matches the BEC limit in a two-component Fermi gas with bare interaction $k_Fa_{23}$, and likewise for large positive detuning with $k_Fa_{13}$.

In Fig.~\ref{fig:becbec_gs}(c) we plot the  out-of-plane and in-plane magnetization $\mathcal{S}_z$ and $\mathcal{S}_x$, respectively, as a function of detuning. We find that the zero crossing of $\mathcal{S}_z$ coincides with the point where the two order parameters are comparable in size, and that $\mathcal{S}_x$ is near unity at this detuning. As also found in the case of a single interaction strength, the detuning for which we have a zero crossing of $\mathcal{S}_z$ is nearly independent of $\Omega$.

Figure~\ref{fig:becbec_gs}(d) shows how the probability of having three particles at close separation remains small throughout the BEC-BEC crossover, peaking at intermediate detunings where there is significant population in both the $\son$ and $\stw$ states. As in the case of a single interaction channel [see Fig.~\ref{fig:thermo1}(b)] this probability is negligible when $|\delta/E_F|\gg 1$ since in these regimes we have an effective two-component Fermi gas.

To further highlight the BEC-BEC crossover, in Fig.~\ref{fig:becbecocc} we show the density distributions and pairing correlations. 
In panels (a)-(c) we plot the density distributions $n_{\k ij}$, which are primarily governed by bound-state wave functions for all detunings. We see that, for large negative detuning in panel (a), pairing is dominated by species $\stw$ and $\sth$, leading to the large occupation of $n_{\k22}$. As the detuning increases into the crossover regime in panel (b), both $n_{\k11}$ and $n_{\k22}$ contribute, signaling a balanced system where pairing contributions come from both scattering channels. Finally, in the large positive detuning limit in panel (c), the density $n_{\k11}$ dominates and the pairing comes from the species $\son$ and $\sth$.

Panels (d-f) illustrate the pairing correlations $\kappa_{\k ij}$, which describe the momentum-dependent structure of the condensate wavefunction across the BEC-BEC crossover. For negative detuning in panel (d), $\kappa_{\k 12}$ is the dominant pairing component, while $|\kappa_{\k 13}|$ remains suppressed. As the system enters the crossover region in panel (e), both $\kappa_{\k 12}$ and $\kappa_{\k 13}$ contribute equally (with opposite sign), demonstrating again that the transition between the two pairing states is smooth.
In panel (f) $\kappa_{\k 13}$ becomes the dominant correlation, indicating that the system has transitioned into a molecular condensate of species $\son$ and $\sth$. The dashed black lines in panels (d-f) illustrate that the results quickly approach the expected large-$k$ asymptotic behavior.

\subsubsection{Excited-state crossover}

We now look at the excited state for the same set of interaction parameters, $k_F(a_{13},a_{23})=\sqrt{2}(0.6,0.8)$. We reiterate that the excited state is topologically distinct from the ground state and corresponds to a saddle point of the free energy---see the states indicated with white stars in Fig.~\ref{fig:FreeE}(a). %

\begin{figure}
\includegraphics[width=0.94
\columnwidth]{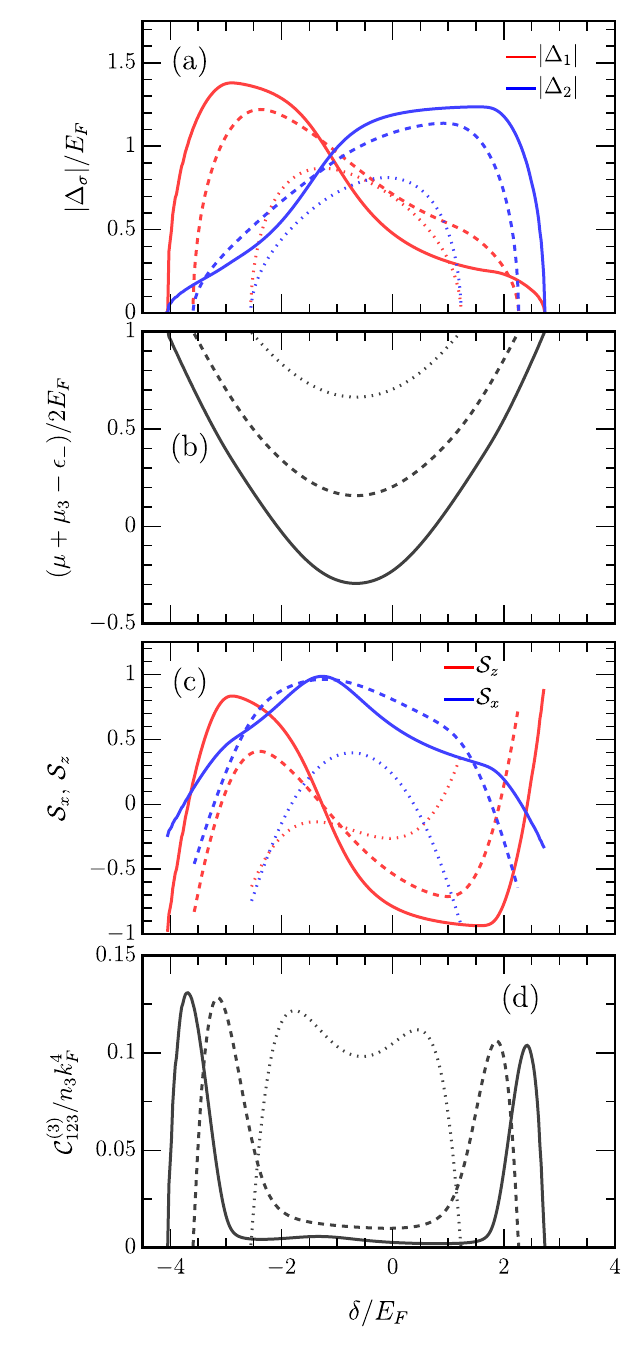}
\caption{Excited branch crossover with Rabi-coupled fermions. (a) Order parameters, (b) total chemical potential, (c) magnetization, and (d) three-body contact. Parameters: $k_F(a_{13},a_{23})=\sqrt{2}(0.6,0.8)$ and $\Omega/E_F=1,2,$ and 3 for the solid, dashed and dotted lines respectively.}
\label{fig:becbec_ex}
\end{figure}

Figure \ref{fig:becbec_ex}(a) shows the order parameters as a function of detuning. We clearly see that the excited state only exists for a limited range of detunings around $\delta=0$. Close to the critical detunings, both order parameters are suppressed and we have a BCS-like state, whereas the order parameters can both be comparable to $E_F$ at intermediate detunings where the state features strong BEC-like correlations. These conclusions are further supported by looking at the total chemical potential in panel (b), which (when measured from the shifted continuum) approaches $2E_F$ at the critical detunings, while it can become negative at intermediate detunings provided the Rabi coupling is not too large (see particularly the $\Omega=E_F$ line). These conclusions highlight that, as a function of detuning, the excited branch features a normal-BCS superfluid phase transition, then a BCS-BEC-BCS superfluid crossover, and then again a BCS-normal transition.

As in the case of the ground state, Fig.~\ref{fig:becbec_ex}(c) shows that the superfluid state features a strong change in magnetization as a function of detuning. However, in this case the variation of $\mathcal{S}_z$ versus detuning is not monotonic, and $\mathcal{S}_z$ can potentially change sign multiple times. Indeed, $\mathcal{S}_z$ crosses zero near the detuning where $|\Delta_1|\simeq|\Delta_2|$ similarly to the ground state, but with an opposite slope and has additional curve turns in the vicinity of the BCS-normal state transitions. The in-plane magnetization $\mathcal{S}_x$ also exhibits an extremum around the detuning where $|\Delta_1|\simeq|\Delta_2|$, but its sign is reversed with respect to that of the ground state. 

In Fig.~\ref{fig:becbec_ex}(d), we plot $\mathcal{C}_{123}$ for the excited state. We see that %
the contact has maxima of the same order as in the ground-state BCS regime in Fig.~\ref{fig:thermo1}(b), and peaks when $\mathcal{S}_z\simeq0$ at positive and negative detuning before the BCS-normal state transitions. For intermediate detunings, the three-body contact becomes smaller, particularly we see that when $\Omega=E_F$, in which case we have an intermediate BEC regime for small negative detunings, %
$\mathcal{C}_{123}$ is strongly suppressed. A similar behavior was also found in the BEC-BEC crossover for the ground state, see Fig.~\ref{fig:becbec_gs}(d).

\begin{figure*}
\includegraphics[width=\linewidth]{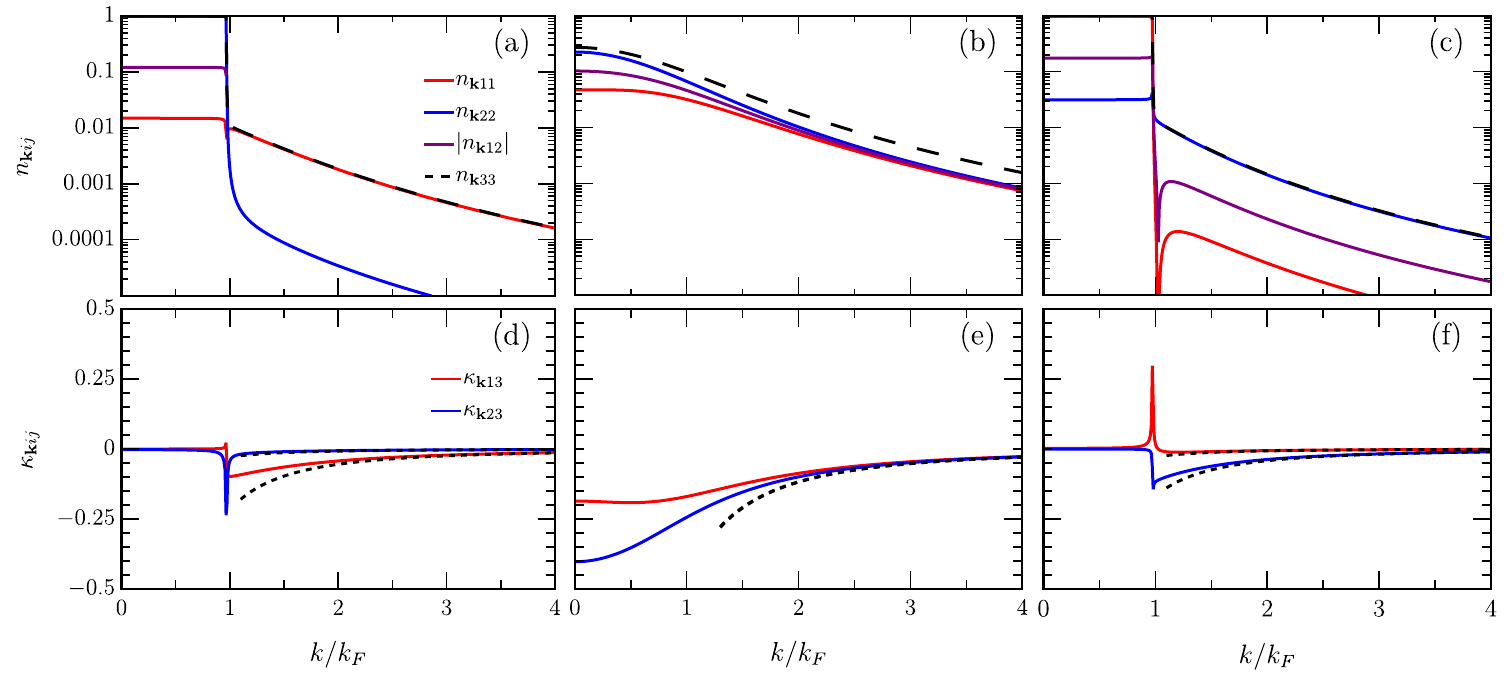}
\caption{(a,b,c) Density distributions and (d,e,f) pairing correlations in the excited branch. From left to right we have $\delta/E_F=-4,-1.1, 2.7$. 
The dashed lines in (d,e,f) show the large-$k$ asymptotes [Eq.~\ref{eq:tailspair}]. 
Parameters: $k_F(a_{13},a_{23})=\sqrt{2}(0.6,0.8)$ and $\Omega/E_F=1$.}
\label{fig:becbecocc_ex}
\end{figure*}

Our conclusions regarding the BCS-BEC-BCS crossover are further supported by the density distributions and pairing correlations shown in Fig.~\ref{fig:becbecocc_ex} for different detunings throughout the crossover. In the BCS regimes (panels a,c,d,f) we see clear signatures of a Fermi sea, whereas in the BEC regime $n_\k$ and $\kappa_\k$ both clearly resemble those obtained in the ground-state BEC-BEC crossover at intermediate detunings shown in Fig.~\ref{fig:becbecocc}(b,e), although in that case the signs of the functions are always opposite.

The excited-state superfluid can potentially be accessed dynamically. The simplest protocol would start with, say, $a_{23}=0$, where we only have a single bound state and a single many-body solution, and then tuning the magnetic field such that $a_{23}>0$. Here, the excited state is continuously connected to the ground state of the initial configuration. The excited state could also play an important role in out-of-equilibrium scenarios, such as those featuring a quench across the superfluid transition~\cite{Gurarie2009,Behrle2018,Dyke2024}.

\section{Conclusion}
\label{sec:conc}

In summary, we have explored the effects of a continuous Rabi drive on a three-component Fermi gas, where two of the components are strongly coupled by the drive. By employing a generalized BCS approach, we derived the free energy, gap equations, and number equations, uncovering a rich phase diagram governed by the parameters of the Rabi drive and by the underlying interparticle interactions. In particular, we demonstrated that the ground state of our system can undergo BCS-BEC, BEC-BEC, and BCS-BCS crossovers by varying the detuning, illustrating how the Rabi-driven Fermi gas is a highly tunable platform for studying many-body physics. We furthermore identified new observables that characterize the crossover, namely the in- and out-of-plane magnetizations associated with the coupled species, and we calculated a disconnected three-body contact parameter that governs the three-body losses. 

We also found the existence of an exotic excited-state solution, which manifests as a saddle point of the free energy. Notably, it exhibits superfluid transitions as well as a BCS-BEC-BCS crossover, and we discussed how these could potentially be accessed dynamically.
Utilizing the free energy, we generalized the Thouless criterion and showed that both the ground and excited states possess critical temperatures that are comparable to those of the standard BCS-BEC crossover, although the excited state critical temperature is somewhat below that of the ground state. Therefore, the physics described in this work should be accessible in current experiments. \bl{To distinguish between the ground and excited branches, two key observables can be used. First, the magnetization of the system, which depends on the detuning, can be reconstructed using spin-resolved imaging, offering a clear signature of the underlying state. 
Second, measuring the pairing gap across the superfluid transition---following the methodology demonstrated in Ref.~\cite{cabrera2024effectstrongconfinementorder}---would provide a direct probe of the many-body state and its evolution with temperature.}

In the future, it would be interesting to investigate beyond-mean-field effects and the role of quantum fluctuations \cite{SadeMelo1993,Ohashi2002,hu2006,Diener2008,Kurkjian2019,powell2022superfluid}, which may significantly impact the stability and nature of the superfluid phases, particularly near the superfluid transition temperature and in the regime where the Rabi-dressed scattering length diverges. 
Additionally, exploring the system in the presence of a spatial magnetic gradient can lead to Fermi gases with position-dependent interactions, modifying the collective excitations and providing a route to engineering a spatial BEC-BCS crossover \cite{spatialBECBCS}.

Another intriguing direction is the dynamics of the system
following a quench or rapid modulation of the detuning, which directly modifies the effective interaction. 
This approach could reveal collective excitations such as Higgs oscillations~\cite{Dyke2024,Kell2024,cabrera2024effectstrongconfinementorder}, as well as Leggett modes~\cite{Leggett66} due to the potential interplay between multiple order parameters.

Finally, it would also be interesting to explore imbalanced densities, i.e., $n_3\neq n_1+n_2$, since there is the prospect of realizing exotic superfluid phases such as the inhomogeneous Fulde-Ferrell-Larkin-Ovchinnikov (FFLO) state \cite{Fulde1964,larkin1965}, where the fermion pairs have non-zero momentum. In an imbalanced two-component Fermi gas, the region of phase space where the FFLO state can be found is expected to be narrow \cite{Sheehy2006,Parish2007,Strinati2018}. Despite considerable effort, experimental observation of the FFLO state remains elusive \cite{Radzihovsky2010}. Controlling the BEC-BCS crossover with the Rabi coupling and detuning offers a promising avenue to explore such exotic phases.

\acknowledgments
We gratefully acknowledge fruitful discussions with Henning Moritz and Nir Navon.
BCM, OB, MMP, and JL acknowledge support from the Australian Research Council (ARC) Centre of Excellence in Future Low-Energy Electronics Technologies (CE170100039), and from ARC Discovery Projects DP240100569 and DP250103746. 
MMP is also supported through an ARC Future Fellowship FT200100619.
CRC is supported by the Cluster of Excellence “CUI: Advanced Imaging of Matter”—EXC 2056—project ID 390715994.

\section{Data Availability}
The data that support the findings of this article are openly available \cite{long_data}.

\appendix

\section{Low-density limit of the BCS ansatz} \label{app:low_dens}

It is instructive to consider the low-density regime of the BCS equations. When the occupations $v_\k^2$ are small, we can take $\theta_\k\ll1$ in which case Eq.~\eqref{eq:minthetaphi13&23} reduces to
 \begin{subequations} \label{eq:minthetaphi13&23-lown_app}
\begin{align} \label{eq:mintheta1323_lown}
& \zeta_\k  \theta_\k +  \left(  c_\k g_{13}\sum_{\q}c_\q \theta_\q +   s_\k g_{23}\sum_{\q}s_\q \theta_\q\right) =0,
\\ \nonumber
& \left[2\delta c_\k s_\k -\Omega  (c_\k^2-s_\k^2)\right]\theta_\k 
\\
& ~~~ -2 \left( s_\k  g_{13}\sum_{\q}c_\q \theta_\q-  c_\k g_{23}\sum_{\q}s_\q \theta_\q \right)  =0 . \label{eq:minphi1323_lown}
\end{align}
 \end{subequations}
Isolating $\theta_\k$ in \eqref{eq:mintheta1323_lown} and injecting it in \eqref{eq:minphi1323_lown}, we obtain
\begin{align} \nonumber
& \left[\delta c_\k s_\k -\frac{\Omega}{2}  (c_\k^2-s_\k^2)\right]\left( c_\k g_{13}\sum_{\q}c_\q \theta_\q+  s_\k  g_{23}\sum_{\q}s_\q \theta_\q \right) 
\\
& ~~~ +\zeta_\k\left( s_\k  g_{13}\sum_{\q}c_\q \theta_\q-  c_\k g_{23}\sum_{\q}s_\q \theta_\q \right)  =0 . 
\end{align}
This equation can be rearranged as a simple equation for $t_\k$
\begin{align} \label{eq:tk1323lown}
t_\k = \frac{\Omega-2(\xi_\k+\xi_{\k3}) R}{2(\xi_\k+\xi_{\k3}+\delta)-\Omega R}, \end{align}
where we have introduced the ratio 
\begin{align} \nonumber
R= -\frac{g_{23} \sum_\q s_\q \theta_\q }{g_{13}\sum_\q c_\q\theta_\q }. \end{align}
We note that $R$ is connected to the ratio of the gap parameters $\Delta_{2}/\Delta_{1}$ when we go beyond the low density limit.

Let us now take a look at Eq.~\eqref{eq:mintheta1323_lown}. Multiplying it by $c_\k$ or $s_\k$ 
one obtains 
\begin{subequations}
\begin{align} 
   c_\k \theta_\k 
&=-\frac{c_\k^2}{\zeta_\k} g_{13}\sum_\q c_\q\theta_\q -\frac{c_\k s_\k}{\zeta_\k}g_{23}\sum_\q s_\q\theta_\q ,\\
  s_\k \theta_\k 
&=-\frac{c_\k s_\k}{\zeta_\k} g_{13}\sum_\q c_\q\theta_\q -\frac{s_\k^2}{\zeta_\k} g_{23}\sum_\q s_\q\theta_\q.
\end{align}
\end{subequations}
Multiplying the first line by $g_{13}$ and the second by  $g_{23}$ and summing over $\k$, these two equations can be rearranged as
\begin{subequations}
\begin{align} 
\frac{1}{g_{13}}
&=-\sum_{\k}\left(\frac{c_\k^2}{\zeta_\k}-\frac{c_\k s_\k}{\zeta_\k}R\right) ,\\
\frac{1}{g_{23}}
&=-\sum_{\k}\left(\frac{s_\k^2}{\zeta_\k}-\frac{c_\k s_\k}{\zeta_\k R}\right) .
\end{align}
\end{subequations}
Using Eq.~\eqref{eq:tk1323lown} to evaluate $c_\k$ and $s_\k$, we obtain
\begin{subequations}
\begin{align} 
\frac{1}{g_{13}}
&=-\sum_{\k}\frac{\xi_\k+\xi_{\k3}+\delta-\frac{\Omega}{2}R }{(\xi_\k+\xi_{\k3})(\xi_\k+\xi_{\k3}+\delta)-\frac{\Omega^2}{4}} ,\\
\frac{1}{g_{23}}
&=-\sum_{\k}\frac{\xi_\k+\xi_{\k3}-\frac{\Omega}{2 R} }{(\xi_\k+\xi_{\k3})(\xi_\k+\xi_{\k3}+\delta)-\frac{\Omega^2}{4}}.
\end{align}
\end{subequations}
Taking $\omega=\mu+\mu_3$, the right hand side can be expressed in terms of  the functions $\Pi_{ij}$
\begin{subequations} \label{eq:tmatstep}
\begin{align} 
\frac{1}{g_{13}}
&=\Pi_{11}(\omega)+R\Pi_{12}(\omega) ,\\
\frac{1}{g_{23}}
&=\Pi_{22}(\omega)+\Pi_{21}(\omega)/R, 
\end{align}
\end{subequations}
where
\begin{subequations}
    \begin{align}
 \Pi_{11}(\omega)&=\sum_\k \frac{c^2}{\omega-\bar{\epsilon}_{\k}-\epsilon_{-}}+ \frac{s^2}{\omega-\bar{\epsilon}_{\k}-\epsilon_{+}},
\\
 \Pi_{22}(\omega)&=\sum_\k \frac{s^2}{\omega-\bar{\epsilon}_{\k}-\epsilon_{-}}+ \frac{c^2}{\omega-\bar{\epsilon}_{\k}-\epsilon_{+}} ,
\\
 \Pi_{12}(\omega)&=\sum_\k \frac{-cs}{\omega-\bar{\epsilon}_{\k}-\epsilon_{-}}+ \frac{cs}{\omega-\bar{\epsilon}_{\k}-\epsilon_{+}}
.
\end{align}
\end{subequations}
Isolating $R$ in the first line of Eq.~\eqref{eq:tmatstep} and injecting in the second we obtain
\begin{align} 
\det[\bold{g}^{-1}-\bold{\Pi}(\omega)]=0,
\end{align}
which corresponds to the equation for the poles of the Rabi-dressed two-body $T$ matrix introduced in Refs.~\cite{mulkerin2024,Shortpaper,Zulli2025}. Thus, similarly to the case of usual BCS theory, the gap equation reduces to the two-body bound state problem in the limit of zero density.

\begin{widetext}
 
\section{Expansion of the free energy for small $\Delta_{i}$}\label{sec:freeEexpansion}

\subsection{Conventional two-component Fermi gas}

It is insightful to remind ourselves how the Thouless criterion arises in the case of the homonuclear two-component ($\up,\down$) Fermi gas with interaction constant $g<0$.
In this case, the mean-field free-energy is given by \cite{Parish2007}
    \begin{align} 
\mathcal{F}_{2c} &= - \frac{\Delta^2}{g} +\sum_\k \left(\frac{\bar{\xi}_\k}{2}-E_\k \right)- \frac{1}{\beta}\sum_{\k}\ln(1+e^{-\beta (E_\k+h)})- \frac{1}{\beta}\sum_{\k}\ln(1+e^{-\beta (E_\k-h)}).  
\end{align}
In this subsection only, we will use the notation $E_\k=\sqrt{\bar{\xi}_\k^2/4+\Delta^2}$, $\bar{\xi}_\k=\xi_{\k\up}+\xi_{\k\dn}$ with $\xi_{\k\sigma}=\epsilon_\k-\mu_\sigma$. In addition, we have introduced $h=(\mu_\up-\mu_\dn)/2$ which plays the role of a Zeeman field which can split the $\up-\dn$ components.
In the limit of small $\Delta$, the free energy can be expanded as
    \begin{align} \nonumber
\mathcal{F}_{2c}&\simeq   \sum_\k\bar{\xi}_\k\Theta(-\bar{\xi}_\k) -\frac{1}{\beta}\sum_{\k}\ln(1+e^{-\beta \left(\frac{|\bar{\xi}_\k|}{2}+h\right)})  -\frac{1}{\beta}\sum_{\k}\ln(1+e^{-\beta \left(\frac{|\bar{\xi}_\k|}{2}-h\right)})  - \Delta^2 T^{\text{mb}}(0)^{-1} + \mathcal{O}(\Delta^4)\\
&=   -\frac{1}{\beta}\sum_{\k \sigma}\ln(1+e^{-\beta \left(\epsilon_\k-\mu_\sigma\right)}) - \Delta^2 T^{\text{mb}}(0)^{-1} + \mathcal{O}(\Delta^4).
\end{align}
We can recognize the free energy of the non-interacting Fermi gas in the first term, while corrections due to the pairing appear in the second. Here, we have introduced the many-body $T$ matrix
\begin{align} \label{eq:TmbappB}
T^{\text{mb}}(0)^{-1}=\frac{1}{g}+\sum_\k\frac{ 1-f_{\k\up}-f_{\k\dn}}{\bar{\xi}_\k}.  
\end{align}
with the Fermi occupations $f_{\k\sigma}=1/(\exp(\beta\xi_{\k\sigma}+1)$. To obtain the above we have used the property of the Fermi-Dirac distribution $f(x)=1/(\exp(\beta x)+1)=1-f(-x)$. As usual, Eq.~\eqref{eq:TmbappB} can be renormalized by replacing $g$ with the $s$-wave scattering length $a_s$ via $\frac{1}{g} = \frac{m}{4\pi a_s} - %
\sum_{\k}^{\Lambda} \frac{1}{2\epsilon_{\k}}$.

It is clear that a second order phase transition can occur when $T^{\text{mb}}(0)^{-1}$ changes its sign, which corresponds to the so-called Thouless criterion \cite{Thouless1960}. In addition, we can see that in the limit of vanishing $\Delta$, the positivity of the quasiparticle energies $E_\k \pm h$ requires $h=0$.
In fact, when $h\neq0$, the system can undergo a first order transition where $\Delta$ exhibits a discontinuous jump \cite{Parish2007}, which is not captured by the above Thouless criterion. 

\subsection{Rabi-coupled three-component case} \label{subsec:freeEexpansion}
We wish to carry out a similar expansion in the case of the three-component Fermi gas with two order parameters, and thus demonstrate Eq.~\eqref{eq:Ftotexpansionfinal} of the main text. Using the fact that the first derivative of the quasiparticle energies vanishes in the limit of zero order parameters, i.e., $\partial E_{\k j}/\partial\Delta_{i}\rvert_{\Delta_{1,2}=0}=0$, we find that the free energy can be expanded to second order as 
 \begin{align} \nonumber
\mathcal{F}_{\text{tot}} &\simeq -\Delta_{1}^2 \underbrace{\left[\frac{1}{g_{13}}+\frac{1}{2}\sum_\k\left(\frac{\partial^2E_\k}{\partial\Delta_{1}^2}  + R \frac{\partial^2E_\k}{\partial\Delta_{1}\partial\Delta_{2}}  \right)+\frac{1}{2} \sum_{\k j}   f_{\k j} \left(\frac{\partial^2E_{\k j}}{\partial\Delta_{1}^2}  + R \frac{\partial^2E_{\k j}}{\partial\Delta_{1}\partial\Delta_{2}}  \right)\right]_{\Delta_{1,2}=0}}_{A}
\\ \nonumber
&~~~-\Delta_{2}^2 \underbrace{\left[\frac{1}{g_{23}}+\frac{1}{2}\sum_\k\left(\frac{\partial^2E_\k}{\partial\Delta_{2}^2}  + R^{-1} \frac{\partial^2E_\k}{\partial\Delta_{1}\partial\Delta_{2}}  \right)+\frac{1}{2}\sum_{\k j}  f_{\k j} \left(\frac{\partial^2E_{\k j}}{\partial\Delta_{2}^2}  + R^{-1} \frac{\partial^2E_{\k j}}{\partial\Delta_{1}\partial\Delta_{2}}  \right)\right]_{\Delta_{1,2}=0}}_{B}\\
&~~ + \underbrace{\sum_{\k} \zeta_\k\Theta(-\zeta_\k)-\frac{1}{\beta}\sum_{\k j}   \ln \left(1+e^{-\beta E_{\k j}|_{\Delta_{1,2}=0}}\right)}_{\mathcal{F}_0} +\text{higher order terms}
.  \label{eq:Ftotexpansion1}
\end{align}
Here, $f_{\k j}$ are the quasiparticle occupations $f_{\k j}=1/(\exp(\beta E_{\k j})+1)$ with the three quasiparticle energies $E_{\k j}$, where $j\in\{\up,+,-\}$. It is understood that these as well as the derivatives between the brackets are evaluated at $\Delta_{1,2}=0$. 
Note that we have introduced $R=\Delta_{2}/\Delta_{1}$ which remains finite in the limit of vanishing order parameters. As usual, this can be renormalized by using the relation~\eqref{eq:renorm}; for brevity, we will not do this explicitly here.

Making use of the relations $2E_{\k}+E_{\k s } = \sum_{j} E_{\k j}$ (see Section~\ref{sec:QPergs}) and $\partial E_{\k s}/\partial\Delta_{i}=0$, the first two lines can be rearranged as 
\begin{subequations}
    \begin{align} 
A&=\frac{1}{g_{13}}+\frac{1}{2}\sum_{\k j}  \left[\left(\frac{1}{2}-f_{\k j} \right)\left(\frac{\partial^2E_{\k j}}{\partial\Delta_{1}^2} + R \frac{\partial^2E_{\k j}}{\partial\Delta_{1}\partial\Delta_{2}} \right)\right]_{\Delta_{1,2}=0},
\\ 
B&=\frac{1}{g_{23}}+\frac{1}{2}\sum_{\k j}  \left[\left(\frac{1}{2}-f_{\k j} \right)\left(\frac{\partial^2E_{\k j}}{\partial\Delta_{2}^2} + R^{-1} \frac{\partial^2E_{\k j}}{\partial\Delta_{1}\partial\Delta_{2}} \right)\right]_{\Delta_{1,2}=0}
.  
\end{align}
\end{subequations}

We now need to evaluate the terms between the brackets in the limit of vanishing order parameters.
Even in this limit, some care is needed when choosing the solution for $t_\k$. We find that we have 
\begin{align}\label{eq:tknoDelta}
t_\k|_{\Delta_{1,2}=0}=\begin{cases}
   \frac{\Omega-2\bar{\xi}_\k R}{2(\bar{\xi}_\k+\delta)-\Omega R},& \bar{\xi}_\k\geq |\epsilon_{-}|,\\
   \sqrt{\frac{\delta^2}{\Omega^2}+1} -\frac{\delta}{\Omega},& \bar{\xi}_\k<|\epsilon_{-}|,
  \end{cases}
\end{align}
i.e., that the appropriate solution depends on whether we are above or below the Rabi-shifted single-particle continuum.
We can see that the solution for $t_\k$ at large $k$ is identical to the one in the low density limit.
Therefore, we know that the ultraviolet divergence of the first terms in the bracket of $A$ and $B$ is exactly canceled upon the normalization of the bare couplings $g_j$. 

The subtlety of the present finite-density configuration is that, at small $k$, there can be a switch to the solution corresponding to the hybridization in the non-interacting limit where $c_\k,s_\k\rightarrow c,s $. This switch can only occur when $\mu+\mu_3-\epsilon_{-}>0$ and thus it disappears in the BEC limit where the total chemical potential is large and negative.
Using \eqref{eq:tknoDelta}, we find that the quasiparticle energies become
\begin{subequations} \label{eq:quasipzerogap}
\begin{align} \label{eq:qp_ni_up}
&E_{\k\up}\rvert_{\Delta_{1,2}=0}=\xi_{\k3}\Theta\left[\bar{\xi}_\k+ \epsilon_{-}\right]-(\xi_{\k}+\epsilon_{-})\Theta\left[-\bar{\xi}_\k+ \epsilon_{-}\right],\\ \label{eq:qp_ni_+}
&E_{\k+}\rvert_{\Delta_{1,2}=0}= (\xi_{\k}+\epsilon_{+})\Theta\left[\bar{\xi}_\k+ \epsilon_{+}\right]-\xi_{\k3}\Theta\left[-\bar{\xi}_\k+ \epsilon_{+}\right],\\ \label{eq:qp_ni_-}
&E_{\k-}\rvert_{\Delta_{1,2}=0}= (\xi_{\k}+\epsilon_{-})\Theta\left[\bar{\xi}_\k+ \epsilon_{-}\right]+(\xi_{\k}+\epsilon_{+})\Theta\left[-\bar{\xi}_\k+ \epsilon_{+}\right]  -\xi_{\k3}\Theta\left[-\bar{\xi}_\k+ \epsilon_{-}\right] \Theta\left[\bar{\xi}_\k+ \epsilon_{+}\right] .
\end{align}
\end{subequations}
These are related to the non-interacting quasiparticle kinetic energies and are independent of $R$, as expected. 
Moreover, the positivity of the quasiparticle energies in Eq.~\eqref{eq:quasipzerogap} imposes constraints on the chemical potentials. Specifically, for equal masses $m=m_3$, we require
\begin{align}
\mu_3-\mu+\epsilon_{-}=0 ~~\text{and}~~~\epsilon_{+}-\mu \geq 0.
\end{align}
Here, the first condition is analogous to the $h=0$ condition in the two component case (i.e., no applied Zeeman field). Combining the first condition with the second gives $\epsilon_{+}-\epsilon_{-}\geq \mu_3$.

The form of the quasiparticle energies above defines three distinct regions of momentum: $\epsilon_{-} \leq \bar{\xi}_\k$, $-\epsilon_{+}<\bar{\xi}_\k<-\epsilon_{-}$ and $\bar{\xi}_\k<-\epsilon_{+}$.
 \begin{table}[t!]
 \caption{\label{tab} Quasiparticle energies and partial derivatives in the limit of vanishing order parameters, in the various regions of momentum.}
\begin{ruledtabular}
\begin{tabular}{llll}
&
(1) ~~$-\epsilon_{-} \leq \bar{\xi}_\k$ & (2) ~~ $-\epsilon_{+}<\bar{\xi}_\k<-\epsilon_{-}$& (3)~~ $\bar{\xi}_\k<-\epsilon_{+}$\\
\colrule
$E_{\k\up}|_{\Delta_{1,2}=0}$  & $\xi_{\k 3}$ & $-(\xi_{\k }+\epsilon_{-})$& $-(\xi_{\k }+\epsilon_{-})$
 \\
$E_{\k+}|_{\Delta_{1,2}=0}$ &  $\xi_{\k }+\epsilon_{+}$ & $\xi_{\k }+\epsilon_{+}$ &  $-\xi_{\k 3}$  
 \\
$E_{\k-}|_{\Delta_{1,2}=0}$   &  $\xi_{\k }+\epsilon_{-}$ &  $-\xi_{\k 3}$ & $\xi_{\k }+\epsilon_{+}$
\\
 $\left(\frac{\partial^2E_{\k \up}}{\partial\Delta_{1}^2} + R \frac{\partial^2E_{\k \up}}{\partial\Delta_{1}\partial\Delta_{2}} \right)_{\Delta_{1,2}=0}$  & $2\frac{c^2-cs R}{\bar{\xi}_\k+\epsilon_{-}} +2\frac{cs R+s^2}{\bar{\xi}_\k+\epsilon_{+}} $& $-2\frac{c^2-cs R}{\bar{\xi}_\k+\epsilon_{-}}$& $-2\frac{c^2-cs R}{\bar{\xi}_\k+\epsilon_{-}}$
 \\
 $\left(\frac{\partial^2E_{\k +}}{\partial\Delta_{1}^2} + R \frac{\partial^2E_{\k +}}{\partial\Delta_{1}\partial\Delta_{2}} \right)_{\Delta_{1,2}=0}$  & $2\frac{s^2+cs R}{\bar{\xi}_\k+\epsilon_{+}} $ & $2\frac{cs R+s^2}{\bar{\xi}_\k+\epsilon_{+}} $& $-2\frac{c^2-cs R}{\bar{\xi}_\k+\epsilon_{-}} -2\frac{s^2+cs R}{\bar{\xi}_\k+\epsilon_{+}} $ 
 \\
   $\left(\frac{\partial^2E_{\k -}}{\partial\Delta_{1}^2} + R \frac{\partial^2E_{\k -}}{\partial\Delta_{1}\partial\Delta_{2}} \right)_{\Delta_{1,2}=0}$  & $2\frac{c^2-cs R}{\bar{\xi}_\k+\epsilon_{-}}$   & $-2\frac{c^2-cs R}{\bar{\xi}_\k+\epsilon_{-}} -2\frac{s^2+cs R}{\bar{\xi}_\k+\epsilon_{+}}$ & $2\frac{s^2+cs R}{\bar{\xi}_\k+\epsilon_{+}} $ 
   \\ 
$\left(\frac{\partial^2E_{\k \up}}{\partial\Delta_{2}^2} + R^{-1}  \frac{\partial^2E_{\k \up}}{\partial\Delta_{1}\partial\Delta_{2}} \right)_{\Delta_{1,2}=0}$  & $2\frac{s^2-cs R}{\bar{\xi}_\k+\epsilon_{-}} +2\frac{c^2+R^{-1}cs }{\bar{\xi}_\k+\epsilon_{+}} $& $-2\frac{s^2-cs R^{-1}}{\bar{\xi}_\k+\epsilon_{-}}$& $-2\frac{s^2-cs R^{-1}}{\bar{\xi}_\k+\epsilon_{-}}$
 \\
 $\left(\frac{\partial^2E_{\k +}}{\partial\Delta_{2}^2} + R^{-1} \frac{\partial^2E_{\k +}}{\partial\Delta_{1}\partial\Delta_{2}} \right)_{\Delta_{1,2}=0}$  & $2\frac{c^2+cs R^{-1}}{\bar{\xi}_\k+\epsilon_{+}} $ & $2\frac{c^2+cs R^{-1}}{\bar{\xi}_\k+\epsilon_{+}} $& $-2\frac{s^2-cs R^{-1}}{\bar{\xi}_\k+\epsilon_{-}} -2\frac{c^2+cs R^{-1}}{\bar{\xi}_\k+\epsilon_{+}} $ 
 \\
   $\left(\frac{\partial^2E_{\k -}}{\partial\Delta_{2}^2} + R^{-1}  \frac{\partial^2E_{\k -}}{\partial\Delta_{1}\partial\Delta_{2}} \right)_{\Delta_{1,2}=0}$  & $2\frac{s^2-cs R^{-1}}{\bar{\xi}_\k+\epsilon_{-}}$   & $-2\frac{s^2-cs R^{-1}}{\bar{\xi}_\k+\epsilon_{-}} -2\frac{c^2+cs R^{-1}}{\bar{\xi}_\k+\epsilon_{+}}$ & $2\frac{c^2+cs R^{-1}}{\bar{\xi}_\k+\epsilon_{+}} $ 
\end{tabular}
\end{ruledtabular}
\end{table}%
Due to the presence of the Fermi distributions in $A$ and $B$, we need to evaluate their integrands in the three regions separately.
The results are summarized in Table~\ref{tab}. We see that, similarly to the quasiparticle energies, the partial derivatives can also be expressed in term of the non-interacting energies and coefficients.
Using the property of the Fermi-Dirac distribution $f(x)=1-f(-x)$, we find
\begin{subequations}
    \begin{align} 
A&=\frac{1}{g_{13}}+\sum_{\k }\Bigg\{\left[1-f(\xi_{\k3})- f(\xi_{\k}+\epsilon_{-})\right] \frac{c^2-cs R}{\bar{\xi}_\k+\epsilon_{-}} 
+
\left[1-f(\xi_{\k3})- f(\xi_{\k}+\epsilon_{+})\right] \frac{s^2+cs R}{\bar{\xi}_\k+\epsilon_{+}} \Bigg\},
\\ 
B&=\frac{1}{g_{23}}+ \sum_{\k }\Bigg\{\left[1-f(\xi_{\k3})- f(\xi_{\k}+\epsilon_{-})\right] \frac{s^2-cs/ R}{\bar{\xi}_\k+\epsilon_{-}} 
+
\left[1-f(\xi_{\k3})- f(\xi_{\k}+\epsilon_{+})\right] \frac{c^2+cs /R}{\bar{\xi}_\k+\epsilon_{+}} \Bigg\}.
\end{align}
\end{subequations}
Moreover, by evaluating the integrands of $\mathcal{F}_0$ in the three regions, we find that it can be rearranged as
    \begin{align} 
\mathcal{F}_0&= -\frac{1}{\beta}\sum_{\k} \ln(1+e^{-\beta\xi_{\k3}}) -\frac{1}{\beta} \sum_{\k ,\pm}\ln(1+e^{-\beta(\xi_\k+\epsilon_{\pm})}) ,
\end{align}
which exactly corresponds to the free energy of the non-interacting Fermi gas.

Inserting the above expressions back into Eq.~\eqref{eq:Ftotexpansion1}, and recalling that $R=\Delta_2/\Delta_1$, we finally obtain Eq.~\eqref{eq:Ftotexpansionfinal} of the main text
\begin{align} \nonumber
\mathcal{F}_{\text{tot}} &\simeq    -\frac{1}{\beta}\sum_{\k} \ln(1+e^{-\beta\xi_{\k3}}) -\frac{1}{\beta} \sum_{\k \pm}\ln(1+e^{-\beta(\xi_\k+\epsilon_{\pm})}) 
 - (\Delta_{1}, \Delta_{2}) \mathbf{T}^{\text{mb}}(0)^{-1} \begin{pmatrix}
    \Delta_{1}\\ \Delta_{2}
\end{pmatrix}. 
\end{align}

\end{widetext}

\section{Analytical calculation of the order parameter in the BCS limit}
\label{app:BCS_limit}

In this appendix, we derive an analytical form of the order parameter at zero temperature in the weakly interacting limit and in the presence of Rabi coupling. Here, we focus on the case with a single interaction channel, taking $a_{23} \rightarrow 0^-$ and $\Delta_1\equiv\Delta$.

First, the gap equation \eqref{eq:gapBCS1323} for a single interaction at zero temperature can be written as
\begin{align} \label{eq:BCSgap_app}
&\frac{m_r}{2 \pi a_{13} } = - \sum_{\mathbf{k}} 
\left[
\frac{c^2_\k}{2 \sqrt{\zeta_\k^2/4 + \Delta^2 c^2_\k}} - \frac{1}{ \bar{\epsilon}_\k}
\right],
\end{align}
where $\zeta_\k=\bar{\xi}_{\k }+\delta s_\k^2-\Omega c_\k s_\k $, $\bar{\xi}_\k\equiv \xi_\k+\xi_{\k3}$, and $c_\k$ and $s_\k$ are found through the solutions of Eq.~\eqref{eq:tk1323}. 
We see that for weak interactions, where $\mu-\epsilon_{-}\simeq \mu_3\simeq E_F$, this integral contains a logarithmic divergence at small $\Delta$ and must be evaluated carefully.

The calculation of Eq.~\eqref{eq:BCSgap_app} in the weakly interacting limit is further complicated by the fact that  care must be taken in finding the solutions for $t_\k$. Specifically, the solutions of $c_\k$ and $s_\k$ in the region $k\geq k_F$ are well approximated by Eq.~\eqref{eq:approxtk1323}, i.e., we have
\begin{subequations}
\label{eq:cksklargek}
\begin{align}
c_\k^2 &= \frac{ (\bar{\xi}_\k + \delta)^2}{
 (\bar{\xi}_\k + \delta)^2 + \Omega^2/4},  \\
s_\k^2 &= \frac{\Omega^2/4}{
 (\bar{\xi}_\k + \delta)^2 + \Omega^2/4},
\end{align}
\end{subequations}
whereas in the region $k<k_F$,  $c_\k$ and $s_\k$ can be approximated by their corresponding noninteracting coefficients, i.e., $c_\k,s_\k\rightarrow c,s$, as discussed in Appendix \ref{subsec:freeEexpansion}. With this in mind we can rewrite Eq.~\eqref{eq:BCSgap_app} into the sum of two integrals for each region,
\begin{align}
\frac{\pi}{k_F a_{13}} =& \underbrace{\int_0^{1} d x \sqrt{x} 
\left\{-
\frac{c^2}{ \sqrt{\zeta_\x^2/4 + |\Delta/E_{F}|^2 c^2}} + \frac{1}{x}
\right\}}_{I_1} \nonumber \\
&+ \underbrace{\int_{1}^\infty d x \sqrt{x} 
\left\{-
\frac{c_\x^2}{ \sqrt{\zeta_\x^2/4 + |\Delta/E_{F}|^2 c_\x^2}} + \frac{1}{ x}
\right\}}_{I_2},
\end{align}
where we have introduced the dimensionless integration variable $ x = \bar{\epsilon}_\k / E_F $ and defined $\zeta_\x=\zeta_\k/E_F$. 

The leading-order divergence of  $I_1$ can be straightforwardly found by following the standard procedure in the non-driven two component case (see, for example Ref.~\cite{Stoof2009b}),
\begin{align}
I_1 \simeq 2  (1 + c^2) - 3 c^2 \ln\left[2\right] + \frac{c^2}{2} \ln\left[c^2 |\Delta/E_F|^2\right] .
\end{align}
In order to find the leading-order logarithmic divergence of $I_2$, we split the integral into two parts:
\begin{align}
I_2 & \simeq
 \underbrace{\int_{1}^\infty d x \left(\sqrt{x}-1\right) 
\left\{-
\frac{2c_\x^2}{\zeta_\x} + \frac{1}{ x}
\right\}}_{I_2^a} \nonumber \\
& + \underbrace{\int_{1}^\infty d x 
\left\{-
\frac{c_\x^2}{\sqrt{\zeta_\x^2/4 + |\Delta/E_{F}|^2 c_\x^2}} + \frac{1}{x}
\right\}}_{I_2^b},
\end{align}
which is only valid in the weakly interacting limit for small $\Delta/E_F$. Here, $I_2^a$ is well behaved for $\Delta=0$, and $I_2^b$ contains the logarithmic divergence.  Rewriting $I_2^a$ in the form
\begin{align}
I_2^a = \int_{1}^\infty d x \left(\sqrt{x}-1\right) \left\{-
\frac{2c^2}{\bar{\xi}_\x+\epsilon_-} - \frac{2s^2}{\bar{\xi}_\x+\epsilon_+} + \frac{1}{x}
\right\},
\end{align}
we can evaluate the integral directly, to give
\begin{align}
I_2^a = & -2c^2\ln[2]- \frac{1}{2}s^2\ln\left[\frac{\epsilon_{+}-\epsilon_{-}}{E_F}\right] 
\nonumber \\ & + 2s^2\sqrt{\frac{\epsilon_{+}-\epsilon_{-}}{2E_F}-1} \, {\rm atan}\left[ \sqrt{\frac{\epsilon_{+}-\epsilon_{-}}{2E_F}-1} \right] .
\end{align}
The logarithmic divergence for small $\Delta/E_F$ is thus contained in the second part, $I_2^b$. Using Eq.~\eqref{eq:cksklargek}, this evaluates to:
\begin{align}
I_2^b \simeq  \frac{1}{2} c^2  \ln\left[c^2|\Delta_{13}/E_{F}|^2\right] + \frac{1}{2} s^2 \ln\left[\frac{\epsilon_+-\epsilon_-}{E_F}\right].
\end{align}
Combining $I_2^s$ and $I_2^b$ with the expression for $I_1$ then yields the divergent behavior of the gap equation Eq.~\eqref{eq:BCSgap_app} at zero temperature. Substituting $I_1$ and $I_2$ into Eq.~\eqref{eq:BCSgap_app} and rearranging gives the compact analytic form of the order parameter in the weak-coupling limit:
\begin{align}
\Delta = \frac{8}{e^{2}|c|}E_{F}\,{\rm exp}\left(\frac{\pi/c^2}{2k_{F}a_{13}}-\frac{\pi^2}{m_rk_F}\frac{s^2}{c^2}I_+\right),
\end{align} 
where  $I_+$ is given by 
\begin{align}
I_{+}&=-\frac{m_r k_F}{\pi^2}\left[1+\sqrt{\frac{\epsilon_{+}-\epsilon_{-}}{2E_F}-1}\arctan\sqrt{\frac{\epsilon_{+}-\epsilon_{-}}{2E_F}-1}\right].
\end{align}

\bibliography{biblio}

\end{document}